\documentclass[]{aa}  
\usepackage{amsmath}
\usepackage{tablefootnote}
\usepackage{bm}
\usepackage{xcolor}
\usepackage[normalem]{ulem}
\usepackage{graphicx}
\usepackage{txfonts}\usepackage[colorlinks,linkcolor=blue,citecolor=blue,linktocpage=true,breaklinks, plainpages=false,urlcolor=blue]{hyperref}
\usepackage{xcolor}

\usepackage{rotating}
\usepackage{sidecap}
\usepackage{subcaption}
\usepackage{url}
\usepackage{soul}
\hypersetup{allcolors=blue}
\usepackage{float}
\usepackage{xcolor}
\usepackage[tracking=true,kerning=true,shrink=50]{microtype}
\usepackage{lipsum}
\usepackage{txfonts}
\usepackage[export]{adjustbox} % For centering the images
\bibpunct{(}{)}{;}{a}{}{,}
\DeclareUnicodeCharacter{00A0}{~}
\usepackage[us,12hr]{datetime}
\usepackage{ulem}

\usepackage{algorithm}
\usepackage{algpseudocode}
\usepackage{sidecap}

\DeclareMathSymbol{\mh}{\mathord}{operators}{`\-}

\def\Ha{\mbox{H\hspace{0.1ex}$\alpha$}}
\def\I2{IRIS$^{2+}$}
\def\ltau{\ensuremath{\log(\tau)}}
\def\vlos{$v_\mathrm{los}$}
\def\vturb{$v_\mathrm{turb}$}
\def\mgii{\ion{Mg}{II}~h\&k}
\def\mg{\ion{Mg}{ii}}
\def\kms{km\,s$^{-1}$}

\begin{document} 

\title{IRIS NUV diagnostics for Ellerman bombs:\\spectral properties, thermodynamics, and formation height}

\author{
I. J. Soler Poquet\inst{1,2}
\and
C. J. D\'{i}az Baso\inst{1, 2}
\and
A. Sainz Dalda\inst{3,4}
\and
L. H. M. Rouppe van der Voort\inst{1,2}
\and
D.~N\'{o}brega-Siverio\inst{1,2,5,6}
\and
R. Joshi\inst{1,2}
}

\institute{
Institute of Theoretical Astrophysics,
University of Oslo, %
P.O. Box 1029 Blindern, N-0315 Oslo, Norway
\and
Rosseland Centre for Solar Physics,
University of Oslo, % 
P.O. Box 1029 Blindern, N-0315 Oslo, Norway
\and
Lockheed Martin Solar \& Astrophysics Laboratory, 3251 Hanover Street, Palo Alto, CA 94304, USA
\and
Search for Extraterrestrial Intelligence (SETI) Institute, 339 Bernardo Ave, Suite 200, Mountain View, CA 94043, USA
\and
Instituto de Astrof\'isica de Canarias, E-38205 La Laguna, Tenerife, Spain
\and
Universidad de La Laguna, Dept. Astrof\'isica, E-38206 La  Laguna, Tenerife, Spain\\
\email{i.j.s.poquet@astro.uio.no}
}

% \date{Received September 15, 1996; accepted March 16, 1997}
%\date{Draft: compiled on \today\ at \currenttime~UT}

\abstract
% CONTEXT:
{
Ellerman bombs (EBs) are observational signatures of small-scale magnetic reconnection, key to understanding  the lower solar atmosphere. While their role in active regions has been widely studied using the \Ha\ line, near-ultraviolet (NUV) spectra routinely observed by the Interface Region Imaging Spectrograph (IRIS) offer a promising alternative for EB identification, enabling large-scale studies.
}
% AIMS:
{
We aim to identify the most important spectral signatures of EBs in the IRIS NUV spectra. With this, we seek to develop a robust criterion for their detection solely using the IRIS NUV spectra. In parallel, we determine the typical atmospheric stratification associated with EBs.
}
% METHODS:
{
We used four coordinated observations between the Swedish 1-m Solar Telescope (SST) and IRIS.
Using the \Ha\ line as a reference, we detected 18 different EBs and studied their associated IRIS NUV spectra.
In addition, we used the \I2\ inversion tool to infer the temperature, line-of-sight velocity, and non-thermal broadening from the EB spectra.
}
% RESULTS:
{
The defining feature of EBs in the IRIS NUV is the enhancement of the wings of the subordinated \mg\ triplet in between the \mgii\ lines. Inversions reveal that these signatures are produced by localized temperature increase of $\Delta T$$\sim$~1650~K around \ltau$=-3.8$. 
Using only the \mg\ triplet signatures, we found a detection criterion that successfully recovered 14 of 18 \Ha-detected EBs.
In addition, the shape of the \mgii\ lines in relation to the \mg\ triplet can serve as a proxy for the EB formation height.
}
{
The NUV spectrum observed by IRIS is a good candidate for detecting EBs, opening the doors to large-scale studies across the extensive IRIS database, removing the dependence on \Ha\ observations.
This capability bridges the gap between small- and large-scale event studies, which was not possible before due to the lack of coordinated observations. 
}


\maketitle

\section{Introduction}\label{sec:introduction}

Ellerman bombs (EBs) are small and transient brightenings ubiquitous in active and flux emergence regions of the Sun, described for the first time by \cite{ellerman_solar_1917}.
These phenomena provide direct observational evidence of the fast-changing dynamics in the solar atmosphere at sub-arcsecond scales.
High-resolution observations have resolved their intricate morphology, resembling the behavior of a flickering flame when observed close to the solar limb \citep{2011ApJ...736...71W}.
EBs typically present sub-arcsecond sizes and lifetimes of the order of minutes \citep{2013ApJ...774...32V,2019A&A...626A...4V}. Because they are usually found in the vicinity of magnetic field concentrations or near polarity inversion lines, this strongly suggests magnetic reconnection as the driving mechanism  \citep{2002ApJ...575..506G, 2007A&A...473..279P, 2024A&A...683A.190R}.
These small-scale reconnection events are related to changes in the dynamics, energetics, and magnetic field configurations of the solar atmosphere.
Therefore, large-scale statistical studies of EBs are essential to characterize and eventually reveal their global impact.

Traditionally, EBs have been observed using ground-based observatories.
Their main signature appears in the \Ha\ line as an enhancement on the wings while the core remains in absorption. This is the main indicator that EBs are sub-canopy events, occurring in between the higher photosphere and the lower chromosphere \citep{2013JPhCS.440a2007R}.
Beyond \Ha, EBs have been also studied through other hydrogen Balmer lines such as \mbox{H\hspace{0.1ex}$\beta$} or by \mbox{H\hspace{0.1ex}$\epsilon$} \citep{2022A&A...664A..72J,2023A&A...677A..52K}; and the  \ion{Ca}{ii}~8542\,\AA\ and \ion{Ca}{ii} H\,\&\,~K lines \citep[see e.g.][]{2006SoPh..235...75S, %socas-navarro-8564,
2007A&A...473..279P, matsumoto_CaH, 2013ApJ...774...32V, 2024A&A...686A.218N}.
Nonetheless, ground-based telescopes are not optimal for large-scale studies due to their limited continuous observing windows and atmospheric degradation effects.

Space-based facilities offer an alternative due to their uninterrupted observations and the absence of the Earth's atmosphere.
Some attempts have been made to develop detection criteria for EBs using the mid-UV 1600\,\AA\ and 1700\,\AA\ continua solely \citep[see e.g.][]{2019A&A...626A...4V} observed by the Solar Dynamics Observatory's  Atmospheric Imaging Assembly \citep[SDO/AIA;][]{2012SoPh..275...17L}. In particular, \cite{2025A&A...699A..54S} employed deep learning techniques to overcome the limitations of traditional methods.
However, the lack of a detailed spectrum in broadband mid-UV imaging did not allow a comprehensive study of EBs using these diagnostics alone.
A more promising window is the near-Ultraviolet (NUV) spectral range ($2782.7-2835.1$~\AA) observed by the Interface Region Imaging Spectrograph \citep[IRIS;][]{2014SoPh..289.2733D}. 
Due to its wide spectral coverage that includes the \mgii\ lines, the IRIS NUV window samples regions from the upper photosphere to the upper chromosphere \citep{1974ApJ...192..769M, 2013ApJ...772...90L, 2013ApJ...778..143P}, being an excellent diagnostic to study dynamic lower-atmosphere events such as EBs \citep{2016ApJ...824...96T}.
The IRIS NUV is centered at and dominated by the \mgii\ resonance lines, located at 2803.5~\AA\ and 2796.4~\AA\ respectively. 

The \mgii\ lines are accompanied by the subordinate \ion{Mg}{II} UV triplet lines observed at 2791.6, 2798.7, and 2798.8~\AA. 
Their formation region is lower than that of \mgii\ lines. Thus, these lines can be exploited in combination with the \mgii\ lines to probe a wide range of different atmospheric heights \citep{2015ApJ...806...14P}.
The second and third lines of the \ion{Mg}{II} UV triplet (2798.7~\AA\ and 2798.8~\AA) are of particular interest to this study because they are always sampled by the IRIS NUV observations, due to their position between the \mgii\ lines. However, the first line of the \ion{Mg}{II} UV triplet is not always observed, so it will not be used in this work. For simplicity, we will refer to the second and third lines of the \ion{Mg}{II} UV triplet as triplet lines.
The triplet lines are weaker than the \mgii\ lines, usually appearing in absorption and turning into emission during localized heating events in the lower atmosphere.
Although being two different lines, they appear blended in IRIS observations due to its spectral resolution, resembling one single line \citep{2015ApJ...806...14P}.

Over the last decade, some studies have investigated the observational and physical properties of EBs based on the NUV spectra. 
On the observational side, the main feature associated with EBs in the literature is the enhancement of the wings of the triplet while the core remains in absorption.
Other features, such as the strong enhancement of the pseudo-continuum bump between the \mgii\ lines and the broadening of the \mgii\ lines, have also been linked to some EBs.
Furthermore, the cores of the \mgii\ lines have been pointed out as worthless to identify EBs since they sample the overlying chromospheric fibrils \citep{2015ApJ...812...11V, 2016ApJ...824...96T, 2019A&A...626A..33H, 2017ApJ...838..101H, 2019ApJ...875L..30C, 2019A&A...627A.101V, ortiz_ellerman_2020}. 
 Establishing a reliable detection criterion using only IRIS data would open the possibility of studying EBs through the extensive IRIS archive. 
On the physical properties side, the thermodynamic properties of EBs have been investigated either by the inversions of spectral data or by numerical simulations and forward modeling.
These studies indicate that EBs are produced by heating events with temperature increase of $\Delta T$ ranging from 400~K to 2500~K, reaching absolute temperatures between 6000~K and 7500~K. These heating events are usually located around the temperature minimum region (TMR)  \citep[see e.g.][]{2014ApJ...792...13H, 2017ApJ...839...22H, 2019A&A...627A.101V,2021ApJ...921...50H}.

Based on these previous studies, the IRIS NUV spectral window emerges as a powerful candidate for identifying and characterizing EBs. 
However, those works focus on the detailed analysis of individual, isolated events, and the results are not sufficient to characterize EBs in a general way.
To reliably detect EBs solely using IRIS data, a comprehensive understanding of their generic IRIS NUV spectral signature is required. 
Regarding the thermodynamic parameters associated with EBs, current literature lacks a statistical consensus on the atmospheric stratification underlying these events.
To address this gap, in this work, we analyze 18 EBs from four different coordinated observations between the Swedish 1-m Solar Telescope \citep[SST;][]{2003SPIE.4853..341S} and IRIS.
%
% We use the \Ha\ line in combination with the \mgii\ spectra\footnote{By \mgii\ spectra, we refer to the spectral range that includes the \mgii\ lines and the lines of the Mg II UV triplet lines located between the former ones.} to confirm that the events we analyze are real EBs.
%
We use the \Ha\ line in combination with the IRIS NUV spectra to confirm that the events we analyze are real EBs.
On the one hand, we study the spectral signatures of all the EBs in the IRIS NUV spectra, correlating them with their corresponding \Ha\ profiles. With all this information, we develop a robust criterion to detect EBs solely using the IRIS NUV spectra.
On the other hand, we infer the general atmospheric parameters associated with EBs, such as temperature enhancements, line-of-sight velocities, and non-thermal broadening from the analysis of the inversion of the \mgii\ lines.

\section{Observations}
% \subsection{Coordinated datasets}

We used four different datasets obtained through coordinated observation campaigns between SST and IRIS.
The combined spectral coverage from both telescopes contains the \Ha\ line from SST and a long list of lines observed by IRIS in the far ultraviolet (FUV) and NUV spectra, although we will only use the latter.
The SST observations were obtained by the CRisp Imaging Spectropolarimeter \citep[CRISP;][] {2008ApJ...689L..69S}. High data quality was further achieved with the aid of the adaptive optics system \citep{2024A&A...685A..32S}. The observations were processed by means of the SST data reduction pipeline \citep[SSTRED;][]{2015A&A...573A..40D, 2021A&A...653A..68L} which employs the Multi-Object Multi-Frame Blind Deconvolution image restoration technique \citep[MOMFBD;][]{2005SoPh..228..191V}.
In order to work with the raster-file format of IRIS, SST data was transformed to level~3 data format, described by \citet{2020A&A...641A.146R}.
With this, the \Ha\ observations are added to the IRIS raster lines, allowing us to extract the \Ha\ and NUV spectra from the same pixel.
Table~\ref{table:observations} summarizes the most relevant parameters for each observation, and Fig.~\ref{fig:all_obs_snapshots} presents sample images from the four different datasets in different spectral lines.
All observations were targeting active regions with a 16-step dense raster.
Information about data access can be found in Appendix~\ref{appendix:observations}.

\section{Methods}

\subsection{Ellerman bombs detection}\label{sec:eb_detection}

%-----------------------------

We detected EBs using the intensity of the blue and red wings of the \Ha\ line.
We defined the wings intensity as the average intensity of three spectral positions at $\pm$1.1, $\pm$1, and $\pm$0.9~\AA\ with respect to the \Ha\ line core.
% $-$50.2, $-$45.6, and $-$41.1~\kms for the blue wing, and at +50.2, +45.6, and +41.1~\kms\ for the red wing. In both cases, the offsets are with respect to the line core of the \Ha\ line.
%
Then, we employed a 4-level intensity criterion. We defined four intensity threshold values: 1.2, 1.3, 1.4, and 1.5 times the reference profile wings intensity. The reference profile was computed as the median of a quiet-Sun (QS) region of the same observation.
Only one wing is required to be above the threshold to select a pixel. In addition, we removed those profiles with the line core intensity higher than or equal to that of the wings. 
With the intensity thresholds, we segregated the pixels of each observation according to their brightness into four different intensity groups.
%

\begin{figure*}
\centering
\includegraphics[width=1\linewidth]{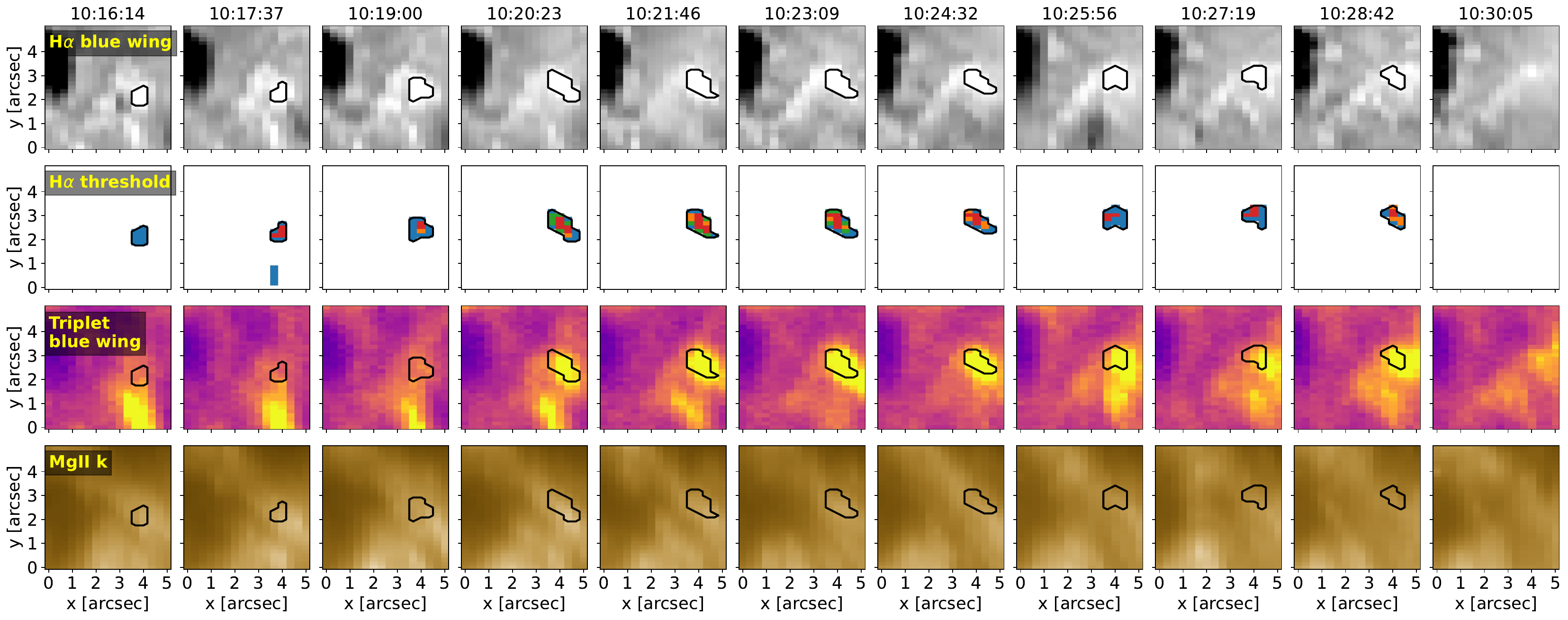}
\caption{ Temporal evolution of an EB in different spectral lines. 
First row: \Ha\ blue wing integration intensity maps.
Second row: Detection mask based on the quadruple intensity threshold method applied to the \Ha\ wings. Each color delineates different intensity enhancements over the reference profile (Blue: $\geq$1.2$\times$, Green: $\geq$1.3$\times$, Orange: $\geq$1.4$\times$, Red: $\geq$1.5$\times$). This color-coding scheme is used throughout the manuscript.
Third row: Triplet blue wing integration (2798.6 -- 2798.8~\AA) intensity maps.
Fourth row: \ion{Mg}{II}~k integration (2795.5 -- 2797.4~\AA) intensity maps.
A contour of the EB based on the \Ha\ mask is drawn on all panels for reference.
All the maps in each row share the same color scale.
% All colormaps are normalized with respect to the entire field of view. 
%
The EB displayed here is EB2 from 2022/06/28 (see Table~\ref{table:eb_information}).
}

\label{fig:EB_multiple_channels}
\end{figure*}

%-----------------------------
%
Then we defined EBs as events consisting of a combination of pixels belonging to the four different intensity groups, with a minimum size of 4 pixels (0.23 arcsec$^2$) and with a minimum lifetime of 3 rasters ($\sim$4 min).
Once the location of the EB is determined, thanks to the \Ha\ line treatment, we can extract its associated IRIS NUV spectrum.
All this process can be seen in Fig.~\ref{fig:EB_multiple_channels}, which presents the temporal evolution of an EB in different spectral lines.
The EB is notable in the \Ha\ blue-wing images (first row), appearing as a bright blob which evolves in intensity and displays some internal structure. This is better seen in the second row, where red pixels are at the center of the EB, surrounded by the other colors.
The triplet row shows a strong intensity enhancement co-spatial and co-temporal to the EB, especially for the middle panels, and a brightness correspondence with the \Ha\ row. 
The \ion{Mg}{II}~k row also shows a slight brightness enhancement, although not as remarkable as the \Ha\ or the triplet ones.

\subsection{Feature selection}\label{sec:features_selection}

The goal of this study is to categorize the principal signatures of EBs in the IRIS NUV, including the \mgii\ lines and the triplet lines.
A possible strategy is to feed machine-learning models with the complete spectral range so they can autonomously find the relevant information \citep[see e.g.,][]{kleint2022occurrence, 2024A&A...689A..72Z, 2021A&A...652A..78S}. However, this approach is not optimal for our particular problem due to two main reasons. 
%first reason
The first one is that EBs co-observed by SST/\Ha\ and IRIS are rare events. To find patterns related to an event and be able to describe it, the number of examples of that event has to be statistically representative in the data \citep[see, e.g.,][]{2025A&A...699A..54S}.
%second reason
The second reason is that a machine-learning model can highlight uninformative features as important and not focus on the meaningful ones due to the scarcity of the EB sample \citep[see, e.g.][]{2020ApJ...891...17P}. Furthermore, extracting the important features from a well-built model is not trivial.
Thus, we decided to reduce the dimensionality of the data to a small number of features, for which we know their physical origin. This allowed us to physically interpret the results.
The selected features are listed below:

\begin{itemize}
%triplet
\item Triplet integration: To get an estimate of the triplet total enhancement, we averaged the interval $2798.600 - 2799.094~\AA$.
\item Triplet blue and red wing, and core intensities: to find their position, we used the extremum-finding algorithm defined by \citet{2013ApJ...772...90L} using the wavelength 2798.8~\AA\ as offset and the interval $2798.600 - 2799.094~\AA$.
%of $\pm$40~\kms$.

%bump between h and k
\item Center of the bump between the \mgii\ lines: The \mgii\ lines are very strong and have extended spectral line wings. These wings overlap between the \mgii\ central emission features and together produce a bump. According to \citet{2013ApJ...778..143P}, this spectral range spanning from  2800.34 to 2800.51~\AA\ relates to the photospheric temperature at heights around 0.28\,Mm. We used the average of the intensity in this interval as a feature.

%Mgh&k
\item \ion{Mg}{II}~k width: To robustly quantify line broadening independent of peak intensity variations, we calculated each profile's normalized cumulative distribution function as done in \citet{2020ApJ...891...17P}. Thus, we extracted the line width as the difference between the third and the first quantile. The interval used was $2795.3-2797.4\,\AA$. 
\item \ion{Mg}{II}~k integration: The line integration was obtained from the same interval as the line width.
\item \ion{Mg}{II} k$_\mathrm{2v}$, k$_\mathrm{2r}$, k$_{3}$: We used the same algorithm as for the triplet wings and core position, with the offset at 2796.4~\AA\ and an interval of $\pm0.37\AA$ ($\pm$40~\kms) with respect to the offset.

\end{itemize}

In addition to the listed features, we also computed the Doppler shift with respect to their positions in the reference profile of the following spectral features: the \ion{Mg}{II}~k$_\mathrm{2v/r}$, k$_\mathrm{3}$, triplet blue and red wing, and core of the triplet.
The locations of the spectral lines mentioned above are indicated in Fig.~\ref{fig:EB1_2024_05_21}.

\subsection{Inversions}

To understand the formation and characteristics of EBs, we need to study their atmospheric properties.
%inversions
Inversion codes infer an optimal model atmosphere that reproduces the observed spectrum. 
Given an initial model atmosphere, the inversion problem is resolved by synthesizing a new emerging spectrum using the radiative transfer equations. 
The synthetic spectrum is compared with the observations, and the atmospheric model is modified to minimize the difference between the synthetic spectrum and the observed spectrum.
This process is done iteratively until a satisfactory fit is obtained. 
For extensive reviews, see \citet{2016LRSP...13....4D, 2017SSRv..210..109D}.
The disadvantage of advanced radiative transfer solvers is their computational cost, increasing as we introduce more complicated physics such as non-local thermodynamic equilibrium (non-LTE) and partial frequency redistribution (PRD). For instance, the inversion of a single \ion{Mg}{II}~h\&k profile under non-LTE and PRD using the Stockholm Inversion Code \citep[STiC;][]{stic} requires approximately 1.5 CPU hours \citep{2024ApJS..271...24S}.
Therefore, performing traditional pixel-by-pixel inversions on large statistical datasets of EBs is computationally prohibitive.

\subsubsection{IRIS$^{2+}$}\label{sec:iris2+}

To provide a statistically robust analysis over thousands of pixels without the prohibitive computational cost of typical non-LTE inversion codes, we employ the \I2 inversion tool \citep{2024ApJS..271...24S, 2026ApJ...997..229S}. This tool represents an optimal balance between capturing complex non-LTE physics and achieving dataset-scale feasibility, as its execution time is faster than standard inversion codes like STiC by a factor of $10^5-10^6$.
The tool is based on a database of 135\,472 synthetic representative profiles (RPs) and their corresponding representative model atmospheres (RMAs) derived from a diverse set of observations. These RMAs provide the thermodynamic stratification of the solar atmosphere from the photosphere to the top of the chromosphere. In this study, we utilize \I2 to recover the depth-dependent profiles of temperature ($T$), line-of-sight velocity (\vlos), and microturbulence (\vturb).
The fitting process in \I2 utilizes a k-nearest neighbor (k-nn) algorithm to find the closest RP in the database and assign the associated RMA to that profile. While the \I2 database includes 6 chromospheric and 6 photospheric lines, we restrict our k-nn search to the \ion{Mg}{II}~h\&k and the triplet lines.
In this work, we used the results from the third inversion cycle of \I2, which used 9, 8, and 8 nodes for the temperature, \vlos, and \vturb\ respectively (default option in \I2). 
Nodes are defined as the locations in the atmosphere where the thermodynamic parameters are computed. For the rest of the atmosphere, the parameters are interpolated.
%IRIS2+ uncertainty 
To compute the uncertainty of the atmospheric models returned by \I2, we use Eq.~42 of \citet{2016LRSP...13....4D}. This formulation derives the uncertainty by propagating the residuals between the observed and synthetic profiles through the response functions of the associated atmospheric model. We only compute the uncertainty at the node locations and interpolate over the rest of the atmosphere, as done in \citet{2021A&A...647A.188D}.
%
% Node are defined as the locations in the atmosphere where the thermodynamic parameters are computed. Thus, a higher number of nodes implies a more detailed atmosphere. Nodes are by default distributed equidistantly along the atmosphere, and the parameters value between nodes are interpolated.
% %
% In this work we use the results from the third inversions cycle of \I2\, which used 9, 8, and 8 nodes for the temperature, \vlos, and \vturb\ respectively. 
%
In Appendix~\ref{appendix:inversion_uncertainty} we show two example inferences performed with \I2 and their computed uncertainties for $T$, \vlos, and \vturb.

\subsection{Correlation between features}

We investigated the correlation between the NUV features, the \Ha\ line intensity, and the inferred atmospheric parameters. 
A challenge in this statistical analysis is the uneven number of EBs across the different datasets. Because EBs from the same dataset tend to be similar regarding their spectral features, if we compute correlations with all the EBs from all datasets together, there will be a bias towards the observations with more EBs.
To mitigate this bias, we computed independent correlation matrices for each dataset. We then averaged these individual matrices to produce the final result shown in Sect.~\ref{sec:correlations}. While this approach ensures equal statistical representation across all observation targets, we note that the correlation operation is not additive.

\section{Results}

We detected a total of 18 EBs across the four datasets. Their physical properties are summarized in Table~\ref{appendix:detected_ebs} in the Appendix.
The median maximum area, of 1.25 arcsec$^2$, is consistent with previous studies \citep[see e.g.][]{2019A&A...626A...4V}. It is important to note that the spatial resolution of the telescope used limits the minimum detectable EB area.
The median lifetime of about 13 min is longer than the median value reported in the literature, but still consistent within the expected distribution. 
Although the study was performed for all the detected EBs, we will discuss in detail two specific events that we consider to be representative enough of the different types of EB spectra.
The two selected EBs represent two boundary cases. The rest of the EBs observed can be placed between them.

\subsection{NUV features}\label{sec:nuvfeatures}
%=================================================================================
\subsubsection{EB 1 2024/05/21}
\begin{figure*}
    \includegraphics[width=0.95\linewidth]{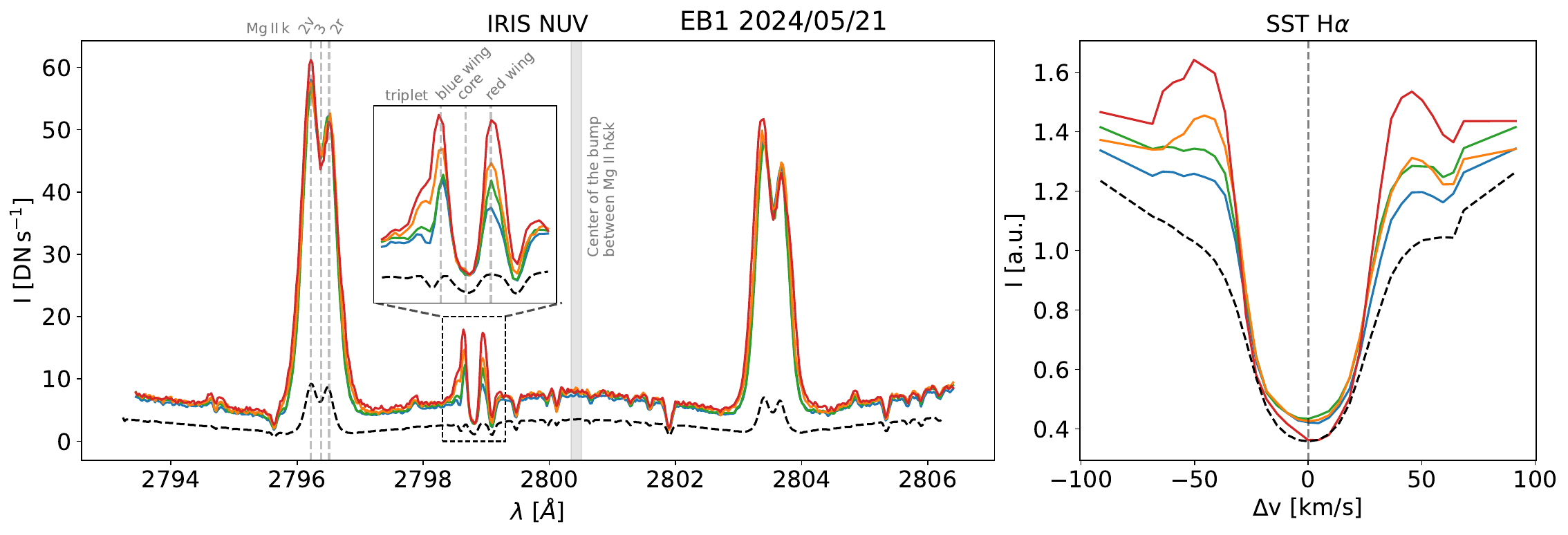}
    \centering
    \caption{IRIS NUV (left) and \Ha\ (right) spectral profiles for EB~1 2024/05/21 segmented by intensity thresholds. Solid lines represent the median spectrum of all the pixels within each intensity group, with the same color-coding as in Fig.~\ref{fig:EB_multiple_channels}. Black dashed spectra represent the quiet-Sun reference profile for comparison.
    Gray dashed vertical lines indicate the spectral position of the \ion{Mg}{II}~k and triplet lines for the quiet-Sun reference profile.
    The locations of the lines used for the analysis defined in Sect.~\ref{sec:features_selection} are noted.
    The inset shows the triplet spectral region in more detail.
    This example shows an EB with strong triplet wings and \mgii\ enhancement, and a relatively weak and flat bump between the \mgii lines.}
    \label{fig:EB1_2024_05_21}
\end{figure*}

Figure~\ref{fig:EB1_2024_05_21} shows the median IRIS NUV profiles (left panel) and the median SST \Ha\ profiles (right panel) for each intensity group (defined in Fig.~\ref{fig:EB_multiple_channels}) for EB1~2024/05/21.
For both lines, we display the median over all the pixels belonging to each intensity group. The left panel shows a general enhancement of the IRIS NUV region for all the profiles with respect to the reference profile.
The outer wings of \ion{Mg}{II}~h\&k, that is, the blue wing of the \ion{Mg}{II} k line and the red wing of the \ion{Mg}{II} h line, and the bump between the \ion{Mg}{II}~h\&k lines exhibit about twice the intensity of the reference profile.
The wings of the triplet present more pronounced increases, with values ranging from 3.7 to 6.5 times the reference intensity.
%, being the red profile the strongest one.
In addition, the wings of the triplet exhibit some degree of broadening, slightly larger for the blue wing. 
The triplet line-core also shows enhancement, although not as strong as the wings one, with an increase of 2.5 times the value of the reference profile.
The \mgii\ lines display considerable enhancement and broadening for all the intensity groups, about 10 times larger than the reference profile.
Unlike the triplet, there is an asymmetry between k$_\mathrm{2v}$ and k$_\mathrm{2r}$, with the former being predominant. The asymmetry indicates a Doppler shift towards the red produced by a non-zero line-of-sight velocity gradient.
This Doppler shift is not seen in the triplet wings, indicating that both lines are formed in different regions.
The right panel in Fig.~\ref{fig:EB1_2024_05_21} presents the SST~\Ha\ profiles. In this panel, the red profile represents a textbook EB profile, where the red and blue wings of the \Ha\ line ($\pm$45~\kms) show a strong enhancement above 1.5 times the reference profile values, which decreases towards the outer wings.

A critical aspect of our analysis is the ordering in intensity of the intensity groups. This allows us to study if the enhancement for the \mgii\ and triplet, and for the \Ha\ is similar or not in EBs, that is, if the brightest pixels in \Ha\ and in the \mgii\ and triplet lines are the same.
For the triplet wings, there is a clear differentiation between the different color profiles. The color sequence (blue-green-orange-red) keeps the same order as for the \Ha\ panel, which holds by definition. However, this ordering is not so noticeable in the rest of the IRIS NUV spectra.
This suggests a relation between the wings of the triplet and the \Ha\ line, since brighter pixels in \Ha\ are also brighter in the triplet and vice versa.

%=================================================================================
\subsubsection{EB 3 2023/07/27}
\begin{figure*}
    \includegraphics[width=1\linewidth]
    {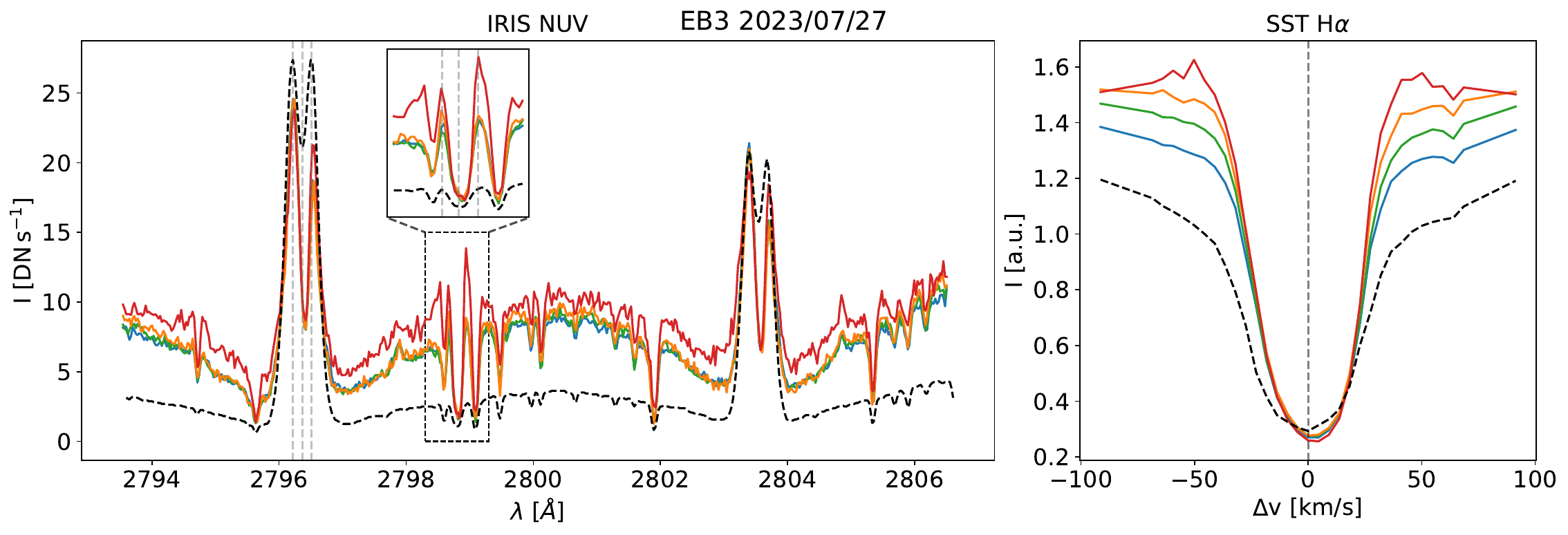}
    \centering
    \caption{Similar to Fig.~\ref{fig:EB1_2024_05_21} but for EB~3 2023/07/27. This example shows an EB with very weak \mgii\ lines, a moderate enhancement of the triplet wings, and a relatively strong and pronounced bump between the \mgii\ lines.}
    \label{fig:EB3_2023_07_27}
\end{figure*}

Figure~\ref{fig:EB3_2023_07_27} shows the features and spectral profiles for EB3~2023/07/27 in the same format as Fig.~\ref{fig:EB1_2024_05_21}. This EB presents a different scenario. 
Because the \mgii\ lines do not exhibit significant broadening, the bump between the \mgii\ lines appears as a prominent convex enhancement (or dome structure).

The intensity of the bump between the \mgii\ lines of the red profile is approximately 3 times larger than the reference one. For the rest of the spectra, the increase is smaller.
The triplet wings enhancements are not as strong, and do not stand out as much with respect to the bump between the \mgii\ as in the EB of Fig.~\ref{fig:EB1_2024_05_21}. The maximum enhancement, of 5 times the reference value, is reached by the red wing of the red profile. However, the rest of the profiles remain in absorption.
The \mgii\ lines appear weaker in intensity than the reference profile. The major intensity decrease occurs at the line cores, with their values around half the reference intensity.
As in the previous example, the \mgii\ peaks have an asymmetry, being the k/h$_\mathrm{2v}$ peaks higher than the k/h$_\mathrm{2r}$ peaks, which is not observed in the triplet wings. Actually, the triplet wings show the opposite case for the red profile. The red wing of the triplet is stronger than the blue wing.
The \Ha\ profiles (right panel of Fig.~\ref{fig:EB3_2023_07_27}) show a general uniform enhancement at the wings except for the red profile, which displays a slight intensity increase around $\pm$50~\kms, but is less noticeable than the one in Fig.~\ref{fig:EB1_2024_05_21}.
The \mgii\ lines are red-shifted with respect to the reference profile, which is also visible in the triplet and \Ha\ line-core. 
The color sequence for the IRIS NUV spectra does not show a clear order. The blue, green, and orange profiles overlap in all the IRIS NUV spectra except for the center of the bump between the \mgii\ and the extremes of the outer wings. The red profile has a higher intensity throughout all the spectra except for the \mgii\ peaks, where it overlaps with the other colors. 

We can differentiate the two profiles presented by the dome/flat structure of the bump between the \mgii, produced by the absence/presence of enhancement of the \mgii\ lines, and by the emission strength of the triplet wings with respect to the center of the bump between the \mgii. More pronounced dome shapes usually appear with less enhanced \mgii\ lines and weaker triplet wings emission with respect to the bump. Nevertheless, cases where the bump appears flatter are often accompanied by much stronger triplet wing emission and \mgii\ broadening and enhancement. 
These differences could be markers of different heights of occurrence for the different EBs.

%=================================================================================
\subsubsection{Summary of Ellerman bombs spectral features}\label{sect:summary_spectral_features}
%=================================================================================

\begin{figure*}
    \includegraphics[width=0.85\linewidth]{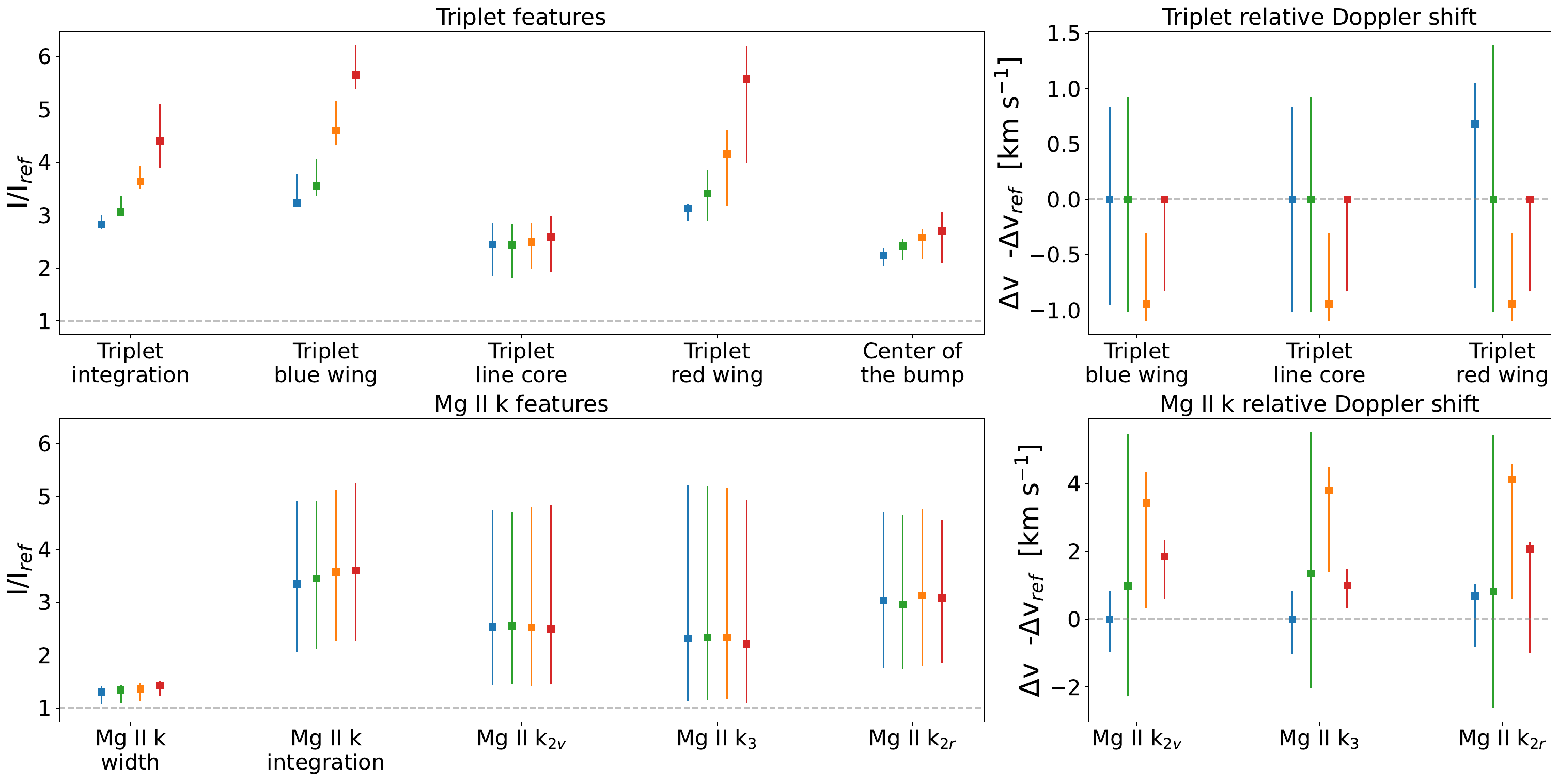}
    \centering
    \caption{Statistical summary of the IRIS NUV spectral features for all 18 detected EBs. Each panel displays the median value of a specific IRIS NUV diagnostic feature across all observations. Values are normalized relative to their respective quiet-Sun reference values from the host observation. Vertical bars indicate the 1$\sigma$ spread (16th to 84th percentiles). Color-coding aligns with the \Ha\ intensity thresholds defined in Fig.~\ref{fig:EB_multiple_channels}.}
    \label{fig:gen_stats}
\end{figure*}

Figure~\ref{fig:gen_stats} summarizes the average value for each spectral feature described in Sect.~\ref{sec:features_selection}, computed over all the EBs of the four observations and segregated by intensity groups.
We normalized each EB feature by the corresponding reference value extracted from its host observation.
% To get each value, we computed each feature value for each EB independently, normalizing by the respective reference feature value. This was extracted from the reference profile of each different observation. 
%
Then, we computed the median for each feature. 
The vertical bar crossing each feature value corresponds to the 16th and 84th percentiles for each intensity group.

The top left panel shows the statistical summary for the triplet features and the center of the bump between the \mgii\ lines.
%Triplet blue wing
The triplet wings display the major enhancement, reaching an increase of 5.5 times the reference value for the red group. This enhancement goes down to 3 times for the blue group.
The triplet integration presents a similar behavior to that of the triplet wings, with enhancements from 2.7 to 4.4 times for the blue and red groups, respectively. 
For the triplet wings and the triplet integration, we can clearly see an ascending color sequence (blue-green-orange-red) similar to the \Ha\ ordering. 
The triplet line core and the center of the bump between the \mgii\ lines present a nearly constant intensity enhancement of around 2.5 times across the different color groups. 
The color ordering for these features is not as pronounced as for the triplet wings and integration, although we can discern some for the bump between the \mgii. 

The lower left panel of Fig.~\ref{fig:gen_stats} displays the statistical summary for the features for the \ion{Mg}{II}~k line.
The \ion{Mg}{II}~k width presents similar enhancement for all the intensity groups of around 1.3, with small dispersion.
The \ion{Mg}{II}~k peaks and integration show a general enhancement around 3 times the reference values. 
Unlike the case of the \ion{Mg}{II}~k width, their dispersion bars are larger or have a comparable size to the median value, indicating that the values of these features are not constrained and are very diverse across the different examples. Another factor is that all the colors share a similar value, which indicates a lack of relation with the intensity as seen in the \Ha\ line.
% Doppler
The upper right panel of Fig.~\ref{fig:gen_stats} shows the relative Doppler velocity for the triplet, and does not show any significant result. All the values are centered or very close to 0~\kms\ with respect to the reference except the orange intensity group, which is centered around $-1$~\kms. 
If we take a look case by case, we can see asymmetries of the triplet wings for 8/18 EBs analyzed.
However, these shifts are different in magnitude and are not present in all cases, 
indicating that the kinematic response is dependent on the specific reconnection geometry, rather than being a defining signature of EBs.
The situation is similar for the \ion{Mg}{II} h\&k Doppler velocity plot (lower right panel), where the values do not follow any particular trend, but an arbitrary order.
The same information as in Fig.~\ref{fig:gen_stats} but in units of DN~s$^{-1}$ is shown in Fig.~\ref{fig:gen_stats_dns} in Appendix D, to facilitate comparison with other observations. 

Examining the spectra of all the detected EBs, we can differentiate three main categories according to the shape of the \mgii\ lines.
The first category comprises profiles where the \mgii\ lines are below the QS reference profile. An example of this is the EB of Fig.~\ref{fig:EB3_2023_07_27}.
The second category are profiles where the \mgii$_\mathrm{2v}$ and h\&k$_\mathrm{2r}$ peaks are enhanced, but the h/k$_3$ peaks are not, remaining similar to the reference profile. An example of this is the EB shown in Fig.~\ref{fig:EB5_2022_06_22}.
The third category includes those \mgii\ lines where all the peaks are enhanced with respect to the reference profile. This is seen in the EB from Fig.~\ref{fig:EB1_2024_05_21}, and is the most represented case in our samples.
The distribution of these categories is not random, but depends on the host active region and not on the individual EB.

The difference in the size of the dispersion and the group ordering between the triplet features and the \ion{Mg}{II}~k features suggests that the triplet emission is physically coupled to the EBs, while the \ion{Mg}{II}~k emission is not, with the exception of the \ion{Mg}{II}~k width.
%

% \begin{figure*}
%     \includegraphics[width=1\linewidth]{figures/EBs_stats_all_obs_QS.pdf}
%     \caption{Statistical summary of the IRIS NUV spectral features for all 18 detected EBs. Each panel displays the median value of a specific IRIS NUV diagnostic feature across all observations. Values are normalized relative to their respective quiet-Sun reference values from the host observation. Vertical bars indicate the 1$\sigma$ spread (16th to 84th percentiles). Color-coding aligns with the \Ha\ intensity thresholds defined in Fig.~\ref{fig:EB_multiple_channels}.}
%     \label{fig:gen_stats}
% \end{figure*}

%=================================================================================
\subsection{Atmospheric stratification}
%=================================================================================

Following the same structure as in the previous section, we will discuss in detail the atmospheric models obtained from \I2 for the same two EBs, followed by the general statistics for all the EB atmospheres.

\subsubsection{EB1 2024/05/21}

\begin{figure*}
    \includegraphics[width=1\linewidth]{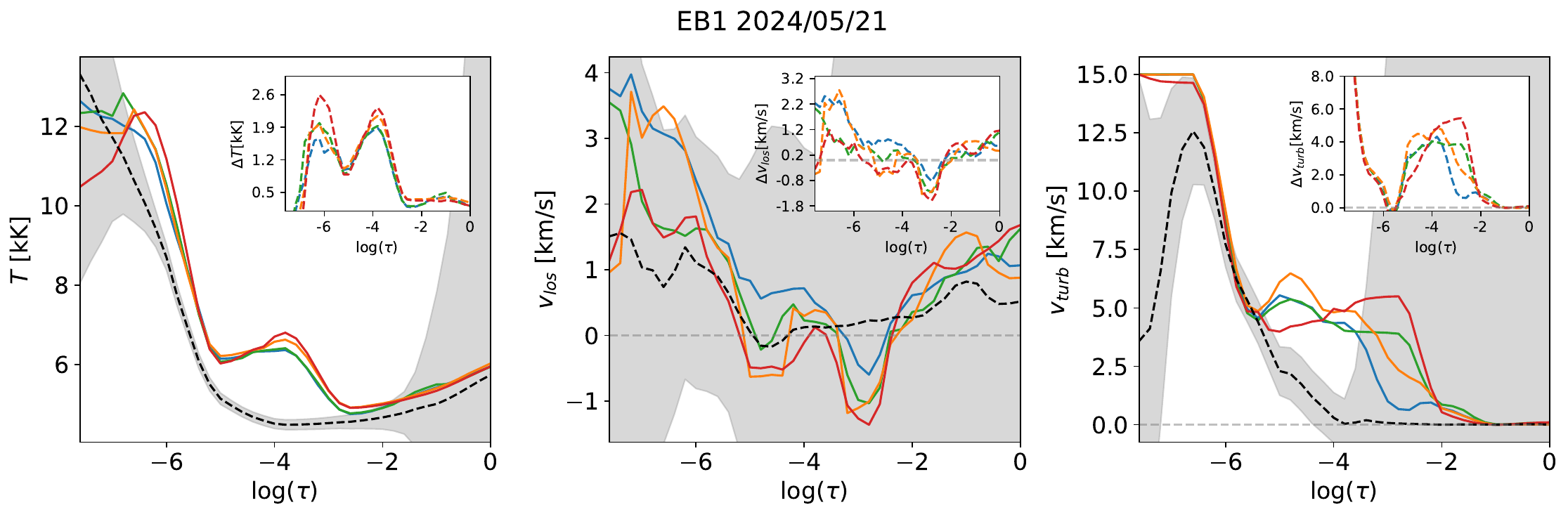}
    \centering
    \caption{Atmospheric stratification inferred from \I2\ for EB1 2024/05/21.
    Left, center, and right panels display temperature, \vlos, and \vturb\ as a function of optical depth, \ltau. Color-coded profiles correspond to the \Ha\ intensity groups from Fig.~\ref{fig:EB_multiple_channels}. 
    Dashed black curves depict the reference profile stratifications.
    Gray-shaded areas around the reference profiles indicate $1\sigma$ uncertainties for the inversion results for each given magnitude based on Fig.~\ref{fig:uncertainty}. Consequently, regions with a thinner shadow correspond to a more reliable result.
    Horizontal dashed gray lines in the velocity panels mark the zero-point. Inset sub-panels display the absolute difference ($\Delta$) between each intensity group and the reference profile.}
    \label{fig:eb1_2024_atm_params}
\end{figure*}

Figure~\ref{fig:eb1_2024_atm_params} shows the median of the atmospheric parameters obtained from \I2 for each color group for EB1~2024/05/21 (same EB as in Fig.~\ref{fig:EB1_2024_05_21}). 
Although the inversions retrieve information across a wide range of optical depths, the spectral lines lack sensitivity in some regions, making the results in those atmospheric heights less reliable. 
To indicate this behavior in Fig.~\ref{fig:eb1_2024_atm_params}, we used gray shadows around the reference profile to indicate $1\sigma$ uncertainties (as defined in Sect.~\ref{sec:iris2+}) for each different physical quantity.
However, these uncertainties are merely indicative, as they have not been calculated specifically for each different EB profile.
Note that all subsequent comparisons, enhancements, and difference values ($\Delta$) are computed with respect to the inverted atmospheric model of the QS reference profile.
The temperature panel shows a general temperature increase of 500~K in the lower atmosphere (around $\ltau=-1.6$), and exceeding 1~kK higher up until $\ltau=-6.6$. 
These different temperature increases are best seen in the inset plot of the first panel.
The most significant feature is the localized, abrupt temperature enhancement, seen as a prominent peak centered around $\ltau=-3.8$.
The major temperature enhancement in this peak corresponds to the red intensity group, with an increase of ~2300~K with respect to the reference temperature, hence reaching a temperature of 6800~K. 
The middle panel of Fig.~\ref{fig:eb1_2024_atm_params} shows the \vlos. 
Although all the atmospheric stratifications fall inside the uncertainty band of the reference stratification, they present similar behaviors between $\ltau=-2$ and $-5$, and develop faster downflows towards the upper atmosphere.
%
% The major departure from the reference stratification is in the higher part of the atmosphere, around $\ltau=-6.6$, where the blue and orange profiles develop an increase in the \vlos, reaching values between 3 and 4~\kms.
%
The right panel of Fig.~\ref{fig:eb1_2024_atm_params} shows the \vturb. We can see a localized increase of the \vturb\ around $\ltau=-4$ with respect to the reference \vturb, reaching values around 5~\kms\ for all the intensity groups. At higher layers, the microturbulence velocity rises again.
In summary, although the entire atmosphere is moderately heated, the most significant phenomenon is the strong localized heating centered at $\ltau=-3.8$, which reaches values from 6300 to 6800~K.
Over the same location, there is also an increase in the \vturb, although there is no clear signature over the \vlos.

\subsubsection{EB 3 2023/07/27}

\begin{figure*}
    \includegraphics[width=0.9\linewidth]{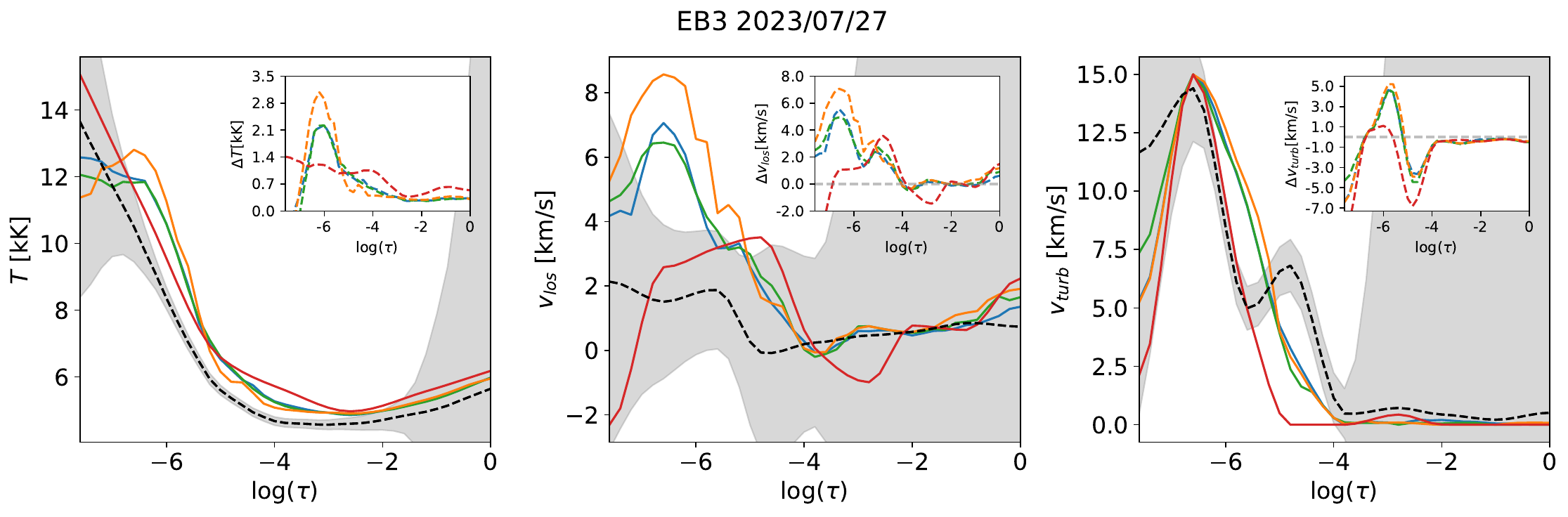}
    \centering 
    \caption{Similar to Fig.~\ref{fig:eb1_2024_atm_params} but for EB3 2023/07/27.}
    \label{fig:eb3_2023_07_27_atm_params}
\end{figure*}

Figure~\ref{fig:eb3_2023_07_27_atm_params} displays the median atmospheric parameters for EB3~2023/07/27 (same EB as in Fig.~\ref{fig:EB3_2023_07_27}) in the same format as Fig.~\ref{fig:eb1_2024_atm_params}.
Unlike in the previous case, there is not a strong temperature increase in the lower atmosphere, but a smooth increase from $\ltau=-2.5$ to $-5$ for the red profile, which is more visible in the inset panel. 
The maximum $\Delta T$ is about 1000~K at $\ltau=-4$, with $T$=5800~K.
In the \vlos\ panel, the atmospheric stratifications show larger \vlos\ increases, starting at $\ltau=-4$ and peaking at $\ltau=-6.6$, with values between 8.5~\kms and 6.5~\kms.
In contrast to the \vlos\ panel of Fig.~\ref{fig:eb1_2024_atm_params}, the velocity values mentioned are partially above the uncertainty.
The right panel of Fig.~\ref{fig:eb3_2023_07_27_atm_params}, which displays the \vturb, does not present any special configuration.
The red profile perceives a step increase, which starts at $\ltau=-4.8$ and peaks at $\ltau=-6.6$. The other color profiles do the same, but starting at $\ltau=-3.8$. All the profiles peak with a value of 15~\kms.

\subsubsection{General EB atmospheric stratification statistics}\label{sec:gen_atm}

\begin{figure*}
    \includegraphics[width=0.9\linewidth]{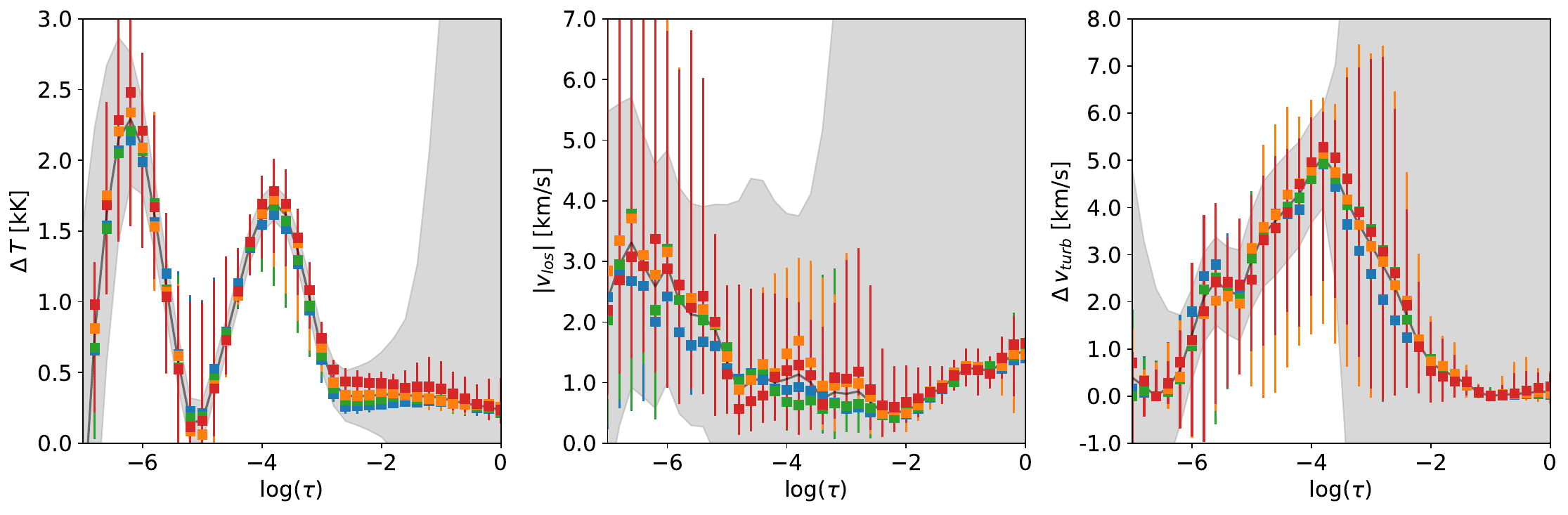}
    \centering
    \caption{Statistical summary of inferred atmospheric parameters for all EBs. Points represent the median value for a given parameter for the different intensity groups. Vertical bars capture variations encompassing the 16th to 84th percentiles. Color-coding is the same as in Fig.~\ref{fig:EB_multiple_channels}. Left: absolute temperature enhancement ($\Delta T$). Middle: absolute line-of-sight velocity ($\Delta$|\vlos|). Right: enhancement in microturbulence velocity ($\Delta$\vturb).
    Gray-shaded areas indicate $1\sigma$ uncertainty with respect to the average stratification of the four intensity groups, represented by the gray solid line.
    }
    \label{fig:ebs_atmos_params_all}
\end{figure*}

Figure~\ref{fig:ebs_atmos_params_all} displays the statistical summary of the atmospheric stratification for all the EBs detected across all the observations. We display the median as well as the 16th and 84th percentiles at each optical depth.
In the left panel, which displays the temperature difference, we can distinguish three different regions with different $\Delta T$ behaviors.
On the lower atmosphere, from $\ltau=-1.75$ to $\ltau=-3$, there is a systematic and constant temperature increase ranging from 260~K to 420~K for the blue and red intensity groups, respectively, with the green and orange in between them. 
In the middle of the atmosphere, there is a very notable temperature enhancement seen as a broad peak extending from approximately $\ltau=-3$ to $\ltau=-5$ for all intensity groups. 
The peak is centered at $\ltau=-3.8$, with maximum $\Delta T$ between 1600 and 1730~K for the different intensity groups, achieving the red profile the largest increment.
The absolute temperatures reached by the different EBs at the center of this peak range from 6100~K to 7000~K. Only three EBs fall below this range (5400~K to 5600~K); however, this is primarily caused by an inadequate \I2\ fit that fails to recover the enhancement in the triplet wings, as will be discussed later.
At the left of the central $\Delta T$ peak, towards the higher atmosphere, there is a step decrease followed by a strong increase. This is probably required by the \I2 to produce a general enhancement of the \mgii\ lines, producing the lowering of the transition region.
The middle panel of Fig.~\ref{fig:ebs_atmos_params_all} shows the absolute values of \vlos, which are non-zero for all \ltau. The value of the \vlos\ increases towards the upper atmosphere, but also the dispersion. However, there is not a particular trend, but a combination of upflows/downflows throughout the atmosphere.
The right panel, which displays the \vturb\ difference, exhibits an almost symmetric increase centered at $\ltau=-3.8$, which equally decreases towards both sides. The peak value for this increase is $\Delta$\vturb=5.3~\kms, which decreases to $\Delta$\vturb=0.3~\kms\ at $\ltau=-6.5$.
There are no noteworthy differences between the different intensity groups. 

\subsection{Features and atmospheric parameters correlations}\label{sec:correlations}

\begin{figure*}
    \includegraphics[width=0.9\linewidth]{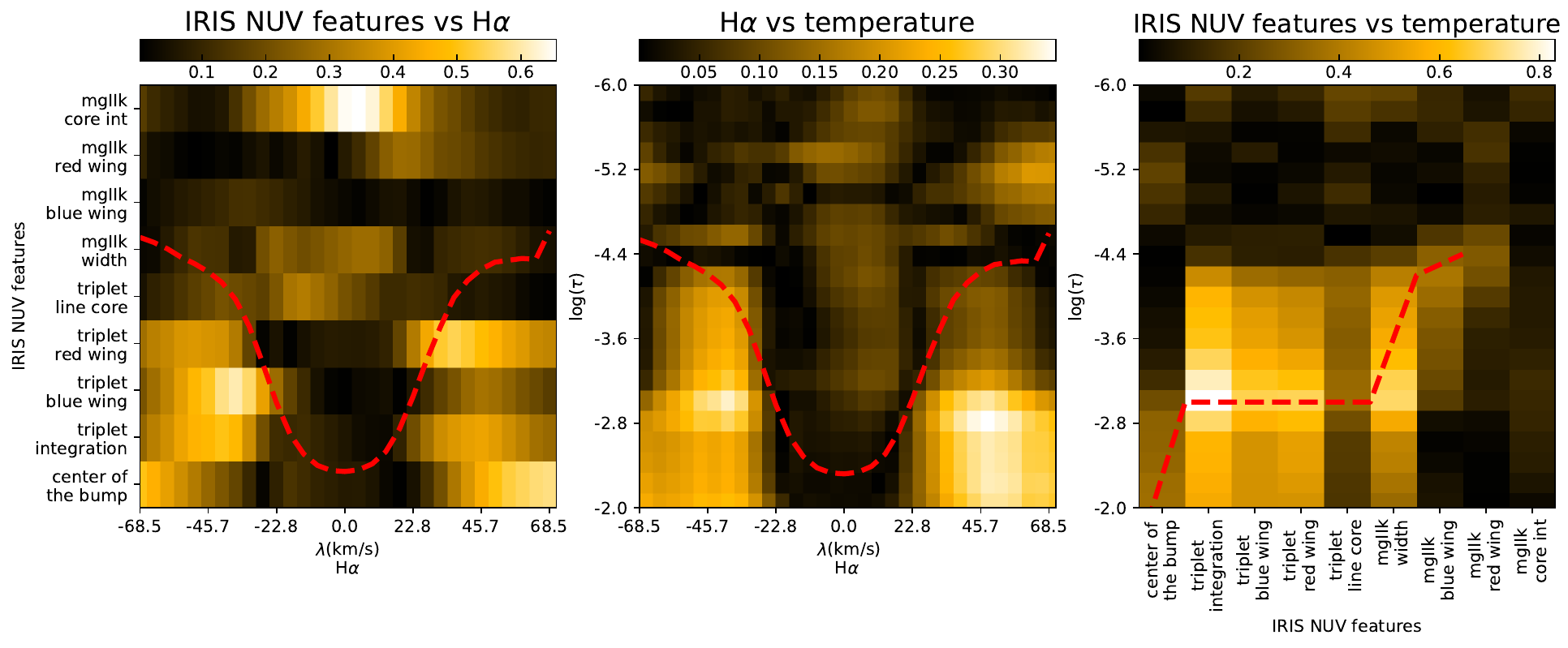}
    \centering
    \caption{Pearson correlation matrices illustrating the statistical relationships between chromospheric spectral features and inferred atmospheric temperatures. {Left:} Correlation between specific spectral positions within the \Ha\ profile and the extracted IRIS NUV spectral features. {Middle:} Correlation between the \Ha\ intensity profile and the inferred atmospheric temperature across different optical depths. {Right:} Correlation between the extracted IRIS NUV spectral features and the atmospheric temperature stratification. The red dashed line shows a typical \Ha\ profile for reference in the left and middle panel, and indicates the \ltau\ of maximum correlation for each feature in the right panel.}
    \label{fig:correlation_matrix}
\end{figure*}

To quantitatively validate the physical connections of the previous sections, we evaluate the statistical correlations between the ground-based \Ha\ spectra, the space-based IRIS NUV features, and the inverted atmospheric temperatures (Fig.~\ref{fig:correlation_matrix}).
The left panel shows the correlation matrix between the intensity for the spectral positions in the \Ha\ line, and the intensity of the features in the \mgii\ lines and the triplet. The maximum correlation is between the \Ha\ line core and the \ion{Mg}{ii}~k$_{3}$, with a value of 0.65. After this, the most correlated parts are the \Ha\ blue wing (around $-40$~\kms) with the triplet blue wing intensity, with a correlation of 0.61, and the \Ha\ red wing (around +40~\kms) with the triplet red wing intensity, with a maximum correlation of 0.54. This is followed by the bump between the \mgii\ lines, whose correlation coefficient increases as we go further in the wings of \Ha. 
This result points to the connection between the wings of the \Ha\ line and the wings of the triplet as complementary diagnostics. 

The middle panel shows the correlations between the temperature at different heights in the atmosphere and the \Ha\ spectral line. The correlations are very weak, with a maximum of 0.35 for the \Ha\ red wing with the temperature at $\ltau=-2.8$. However, we can see patterns which indicate that the temperatures in the interval $\ltau=[-1.2,-3]$ are more correlated with the \Ha\ wing intensities than temperatures at other heights.
In contrast, the right panel reveals the strong correlations between the IRIS NUV features and the temperature stratification.
A highly correlated region appears between the triplet integration and the \ion{Mg}{II}~k width, and it is centered around $\ltau=-3$. At this optical depth, the integrated triplet intensity achieves a maximum correlation of 0.83. This is followed by the \ion{Mg}{II}~k width, the triplet blue, and the triplet red wing, with maximum values of 0.70, 0.69, and 0.69, respectively, all for $\ltau=-3$. 
The center of the bump between the \mgii\ lines, however, presents its maximum correlation at lower heights, around $\ltau=-2$.
The correlation at different heights indicates that the strength of the features can be used to infer the height formation of the EBs.

\subsection{Ellerman bombs formation height estimation}

To quantitatively verify our hypothesis that the IRIS NUV spectral morphology encodes the formation height of EBs, we computed the response function of a representative EB profile.
The response function indicates how sensitive a spectral line is to a local perturbation of a given physical quantity in a given atmospheric height.
In particular, we studied how the strong EB profile used for the uncertainty calculation (Fig.~\ref{fig:uncertainty}) varies in response to perturbations in the temperature stratification.

\begin{figure*}[t]
\begin{minipage}[b]{0.7\textwidth}
\includegraphics[width=\linewidth]{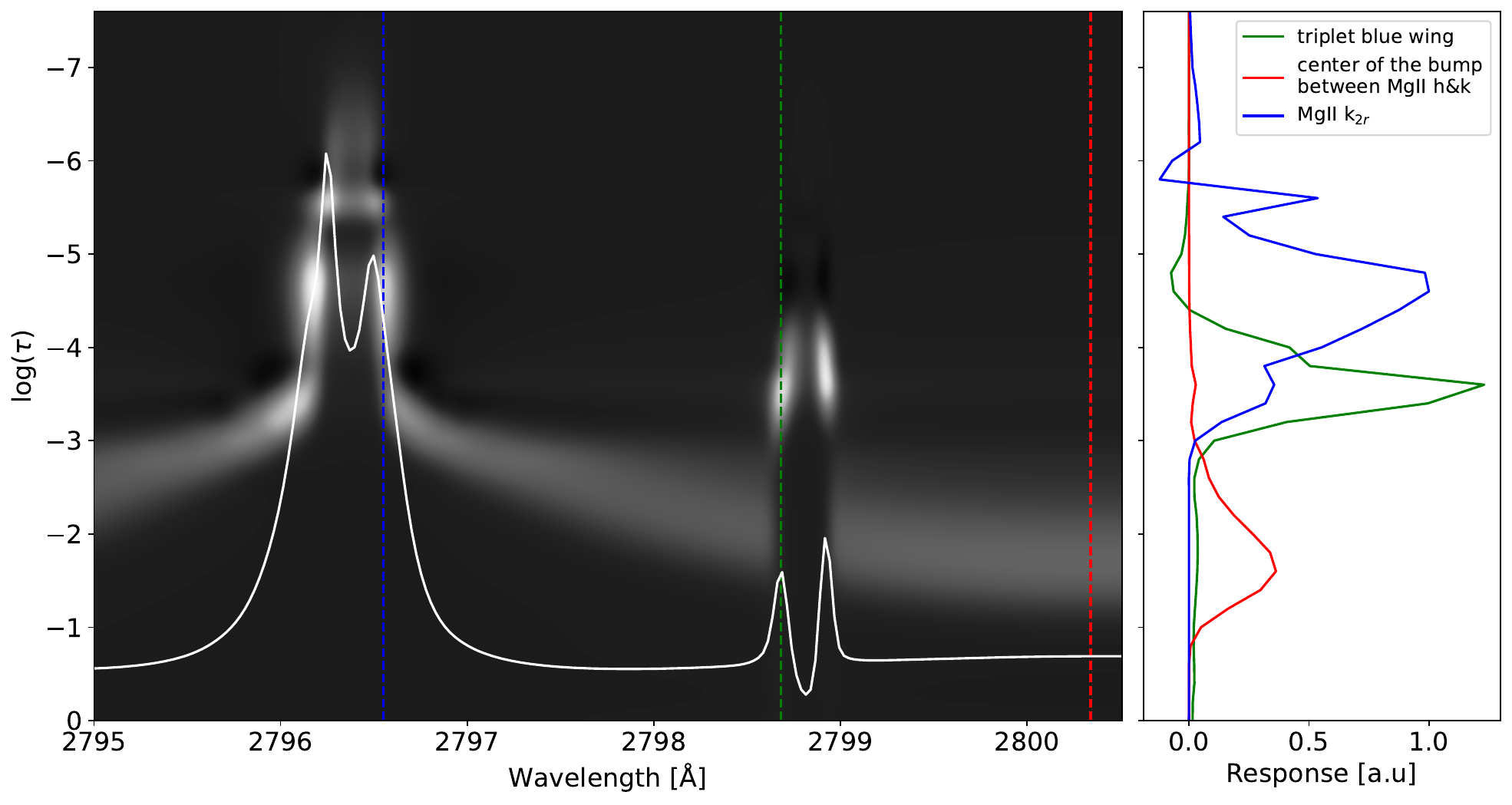}
\end{minipage}%
\hspace{\fill}%
\begin{minipage}[b]{0.25\linewidth}
\caption{Left: response function of the intensity to perturbations in temperature of a strong EB profile (same one as in Fig.~\ref{fig:uncertainty}). The corresponding synthesized spectrum is also overplotted for reference (solid white line). Right: response evaluated at the three wavelengths marked by the color lines in the left panel. The blue, green, and red colors correspond to the \ion{Mg}{II}~k$_{2r}$, the blue wing of the triplet, and the center of the bump between the \mgii\ lines, respectively. We omitted the \ion{Mg}{II}~h region since its response is similar to that of \ion{Mg}{II}~k.}
\label{fig:response function}
\end{minipage}
\end{figure*}

The result is shown in Fig.~\ref{fig:response function}. 
The response function, represented by the color map in the left panel, is normalized by the maximum value in it. Then, a higher value indicates that the line is more sensitive to a change in the local conditions.

The different parts of the IRIS NUV spectrum are sensitive to different heights in the atmosphere. 
The region of the bump between the \mgii\ lines, around the red dashed vertical line, forms deeper in the atmosphere, ranging from $\ltau=-0.5$ to $\ltau=-3$, with the peak at \ltau$\sim-1.6$. This value remains constant along the bump.
The responses of the wings of the triplet and of the \ion{Mg}{II}~k$_2$, however, increase drastically in height as we approach the line cores.
In the case of the triplet wings, the response partially overlaps with the bump's response. However, the maximum sensitivity is reached between \ltau$\sim-3$ and \ltau$\sim-4.3$ for the spectral position indicated by the green dashed vertical line. 
The \ion{Mg}{ii}~k$_2$ response covers a wider range than that of the triplet, with heights from \ltau$\sim-2.6$ to $\ltau\sim-5.4$ depending on the wavelength, showing clear overlap with the triplet. 
The partial overlap between the three spectral features analyzed is clearly seen in the right panel of Fig.~\ref{fig:response function}. Thus, depending on where the heating is localized, some features will be enhanced with respect to the others, and can infer information about the formation height.
Stronger triplet wings enhancement combined with broad and enhanced \ion{Mg}{II}~k$_2$ peaks (see e.g.~Fig.~\ref{fig:EB1_2024_05_21}) would indicate the formation of EB higher in the chromosphere than profiles with weaker triplet emission, non \mgii\ enhancement, but more pronounced bump (see e.g.~Fig.~\ref{fig:EB3_2023_07_27}).

\subsection{EB detection with IRIS NUV}

To develop an EB detection algorithm that operates solely on IRIS NUV data, we must overcome the limitations of the relative normalizations used in Fig.~\ref{fig:gen_stats}. 
Those results are produced using a QS reference region for each observation based on the \Ha\ observations.
To remove this dependence on ground-based data, we transition to absolute radiometric units, utilizing the typical EB feature values in data numbers per second (DN~s$^{-1}$) shown in Fig.~\ref{fig:gen_stats_dns}.
Based on these absolute intensities, we found a simple and optimal recipe to detect NUV EBs using the triplet features (computed following Sect.~\ref{sec:features_selection} indications).
First, we classified all the pixels of the observation into two different groups based on the absolute intensity thresholds defined in Table~\ref{table: eb detection}.
For each different feature, particular intensity conditions are indicated. 
If a pixel meets the conditions for the blue group, it is segregated to that group. If the pixel meets the conditions for both groups, the pixel is classified into the red group. However, if the pixel does not meet the conditions, it is discarded. Red and blue pixels represent the strongest and weakest events, respectively, as done for the \Ha\ detections. 
Second, we searched for events with a minimum spatial overlap in time of 1 pixel, with a minimum lifetime of 2.8~minutes (two frames in our data).
Additionally, we impose that each event must contains both red and blue pixels, with a minimum requirement of two red pixels per event. 
Events that meet these requirements are considered to be associated with EBs.
While this baseline recipe proves highly effective, a detailed study of value combinations is out of the scope of this work. 

Applying these criteria, we successfully identify 14 events in clear correspondence with 14 of the 18 ground-truth \Ha\ EBs, with only one false positive.
The detected events do not align perfectly in time and shape with the \Ha\ EBs. The triplet and the \Ha\, although similar in some aspects, form under different conditions, so we cannot expect a perfect morphological correlation. This behavior can be seen in Fig.~\ref{fig:EB_multiple_channels}. \Ha\ and the triplet enhancements are clearly correlated, although they are not exactly the same; they exhibit minor temporal and spatial offsets.
%
% Thus, what we detect are events associated with EBs (or NUV EBs).
%
Consequently, we define the phenomena isolated by this algorithm not as exact morphological clones of \Ha\ EBs, but as their direct upper-chromospheric counterparts: IRIS NUV EBs.

\begin{table}[]
\caption{Summary of threshold values used to detect EBs using triplet line features.}
\begin{tabular}{ccc}
\hline
\begin{tabular}[c]{@{}c@{}}Features  \end{tabular}           & \multicolumn{1}{c}{Blue pixels}                       & Red pixels                         \\ \hline\hline
\begin{tabular}[c]{@{}c@{}}triplet \\ blue wing \end{tabular}  & \multicolumn{1}{c}{9.27 \textless\ $I$ \textless\ 18.72} & 15.78 \textless\ $I$ \textless\ 18.72 \\ \hline
\begin{tabular}[c]{@{}c@{}}triplet \\ red wing\end{tabular}   & \multicolumn{1}{c}{9.01 \textless\ $I$ \textless\ 20.04} & 12.63 \textless\ $I$ \textless\ 20.04  \\ \hline
\begin{tabular}[c]{@{}c@{}}triplet core \end{tabular}       & \multicolumn{1}{c}{$I$ \textless\ 2.99} & \multicolumn{1}{c}{$I$ \textless\ 2.99} \\ \hline
\begin{tabular}[c]{@{}c@{}}triplet \\ integration \end{tabular}& \multicolumn{1}{c}{5.53 \textless\ $I$ \textless\ 10.83} &  8.47 \textless\ $I$ \textless\ 10.83  \\ \hline
\begin{tabular}[c]{@{}c@{}}center of the bump\\ between  \mgii\ \end{tabular}      & $I$ \textgreater 8.06    & $I$ \textgreater 8.06                                    \\ \hline
\end{tabular}
\tablefoot{
For a given pixel, $I$ represents the value of each given feature. The triplet blue and red wings, and core intensities have units of $\mathrm{DN\,s^{-1}}$. The triplet integration and center of the bump have units of $\mathrm{DN\,s^{-1}\,wav^{-1}}$.}
\label{table: eb detection}
\end{table}

\section{Discussion}

\subsection{IRIS NUV spectral signatures for EBs}

%triplet wings
The main spectral signature of EBs in the IRIS NUV is the enhancement of the wings of the triplet line, which is present in all 18 EBs analyzed. The bump between \mgii\ and the triplet core is also systematically enhanced, although it is not as defining as the triplet wings.
These conclusions are extracted from Fig.~\ref{fig:gen_stats}. 
There was one single case in the literature, in \citet{ortiz_ellerman_2020}, where the authors claimed they found an \Ha\ EB example where only a marginal increase of the triplet wings was perceived. 
We re-analyzed that same data, and we found co-temporal significant triplet wings enhancement in the immediate vicinity of the EB in \Ha.
%$(x,y) = (-158.4, 3.7)$
Therefore, our findings are in total agreement with the existing literature \citep[see e.g.][]{2015ApJ...812...11V, 2017ApJ...839...22H, ortiz_ellerman_2020}.

The \mgii\ lines associated with the detected EBs, however, do not present any particular behavior through all the examples, where the \mgii\ lines' shape varies from case to case.
However, the majority of events found in the literature fit in the second trend described in Sect.~\ref{sect:summary_spectral_features}, where the \ion{Mg}{II}~h\&k 2$_{v/r}$ peaks appear enhanced with respect to the QS reference profile, while the h\&k~3 remains unaffected \citep[see e.g.][]{2015ApJ...812...11V}.
This classification depends on the used reference profile, so it is only a guideline. 
In some observations where IRIS is pointing at an active region, it can be challenging to find a proper QS region to use as a reference.

\subsection{Temperature required for EBs}
%Atmospheric parameters

According to our results, the emission of the triplet wings is produced by localized temperature enhancement between 1600~K and 1740~K around $\ltau=-3.8$, with an extension $\Delta$\ltau=2. This sets peak temperatures ranging from $T\sim6100$~K to 7000~K.
The relationship between the enhancement of the wings of the triplet line and the increase in temperature in the lower atmosphere is also supported by the right panel of Fig.~\ref{fig:correlation_matrix}. The clear high correlation zone covers the triplet integration, blue, and red wings. This is in accordance with the description given by \citet{2015ApJ...806...14P} to explain the formation of similar spectral profiles in their simulations and observations.
A work where the authors also inverted EBs taking into account the IRIS NUV was the one presented by \citet{2019A&A...627A.101V}, which is directly comparable with our study. 
They analyzed EBs co-observed with SST and IRIS. However, in addition to the \mgii\ line, they also included the \ion{Ca}{II}~8542\,\AA\ and \ion{Ca}{ii}~H lines.
They found peak temperatures of the order of 6500~K, with a larger vertical extension of $\Delta$\ltau=2.5.
We think this can be produced by the addition of the \ion{Ca}{II}~lines, which are more sensitive to lower heights.
From the forward modeling perspective, \cite{2021ApJ...921...50H} and \citet{2016A&A...593A..32G} reproduced synthetic NUV EB spectra, with localized heating regions with $\Delta T$$\sim$1150 to 3250~K around the TMR. Similar results were also obtained by \cite{2017ApJ...839...22H} through 3D radiative magnetohydrodynamic simulations, reaching peak temperatures of 7500~K. 
Therefore, our results agree with the atmospheric temperature stratifications derived in previous works, and demonstrate the reliability of \I2.

\subsubsection{Estimating EB height via the NUV}

So far, we have focused on the triplet line as the prime diagnostic for EBs. However, we can also infer information from the \mgii\ lines and the bump between the \mgii\ lines in correlation to the triplet.
The relation between the mentioned spectral features is clearly seen in the response function shown in Fig.~\ref{fig:response function}, and also in the right panel of Fig.~\ref{fig:correlation_matrix}, although less obvious.
The partial overlap of the response function between the wings of the triplet, the \ion{Mg}{II}~k$_2$ peaks, and the bump between \mgii\ allows us to use them as a diagnostic for the formation height of EBs.
The dependence between the IRIS NUV shape and the heating height has also been studied by \cite{2016A&A...593A..32G} and \cite{2021ApJ...921...50H}, presenting similar results to ours. They showed how heating deeper in the atmosphere, below the TMR, produces enhancement of the triplet wings and of the bump between the \mgii\ lines, with no notable \mgii\ broadening, thus resembling a dome shape of the bump. But higher heating, slightly above the TMR, produce both enhancement of the triplet wings and broadening and enhancement of the \mgii.
We can conclude that the shape of IRIS NUV is a good indicator of the height of the EB. Deeper heating will produce enhancement of the wings of the triplet line and of the bump between \mgii. Due to the absence of the broadening of the \mgii, the bump will resemble a dome shape. As the location where the heating happens becomes higher in the atmosphere, the enhancement of the triplet wings will also increase, and the bump will lose power. At the same time, the size of the \mgii\ will increase. This increase will produce the broadening of the line, which will hide the dome shape of the bump. 
The two EBs discussed in Sect.~\ref{sec:nuvfeatures} were selected as good representatives for each case.

\subsubsection{Connection to quiet-Sun EBs}

A very interesting discussion is the connection between the counterpart of EBs in quiet-Sun, namely quiet-Sun EBs (QSEBs), and our results.
The QSEB example shown in \citet{2017ApJ...845...16N} with IRIS spectra presents clear enhancement on the triplet wings, a dome-shaped bump between \mgii, and some enhancement on the \mgii~2$_{v/r}$ peaks, keeping the core in absorption.
Three QSEBs presented by \citet{2024A&A...689A.156B} had some enhancement in the wings of the triplet and a clear dome structure. The enhancement produced in the \mgii\ for some of the cases was associated with a heating higher in the atmosphere.
%
% As explained above, deeper heatings in the atmosphere would be indicated by a more pronounced dome shape of the bump between \mgii, with some enhancement in the triplet wings that would diminish as the heating is moved downwards. 
%
Following the explanation for the shape of the IRIS NUV spectra given above, one main difference between QSEBs and EBs would be the height of formation, with the former originating deeper on average. However, further work would be required to prove this.

The deeper origin for QSEBs was first proposed by \citet{2016A&A...592A.100R}, who made the argumentation based on the presence of emission in \Ha\ but absence in \ion{Ca}{II}~8542 for QSEBs. 

\subsubsection{Line-of-sight velocity}
In the literature, chromospheric jets have been associated with EBs marked as strong upflows and downflows produced by the magnetic reconnection process, with velocities up to 25~\kms\ \citep{2017ApJ...839...22H, 2019A&A...627A.101V}.
In our study, we did not find any indication of the periodic appearance of jets together with the EBs. 
An important factor to understand these results is the different contribution of the temperature, \vlos, and \vturb\ to the spectral profile, where the former dominates over the others.
This can be seen in Fig.~\ref{fig:uncertainty}, where the temperature has much lower relative uncertainty than the \vlos. Then, the line is much more sensitive to a change in temperature than a change in \vlos. 
This explains the long dispersion bars for the \vlos\ and the \vturb\ in Fig.~\ref{fig:ebs_atmos_params_all}, which are greater than their central value.
As the temperature contribution to the formation of the line is much higher than the others, we can get similar spectra for a similar temperature but very different \vlos\ and \vturb, constraining the temperature much more.
Therefore, the temperature dominance over the \vlos\ for the line formation together with the statistical summary can hide velocity information.

In addition, \cite{2026ApJ...997..229S} showed that \I2\ tends to underestimate the \vlos\ in comparison with direct inversions made with STiC.
To properly study jet formation in EBs sites, particular studies for each case are required, including an individual inversion for each EB.

\subsection{\I2\ limitations}
%IRIS2 limitations
It is important to mention the limitations of \I2. Since it works as a look-up table, we are limited to the profiles of the database.
We consider that the variety of profiles is good enough to obtain a representative trend for the EBs atmospheric parameters. This can be seen in the dispersion bars obtained and shown in Fig.~\ref{fig:gen_stats} and~\ref{fig:eb1_2024_atm_params}. Another positive proof is the compatibility of the results with the literature and the strong trends seen in the first panel of Fig.~\ref{fig:eb1_2024_atm_params}.
However, the shape of the spectral profiles utilized can produce some non-realistic results.
As mentioned in Sect.~\ref{sec:iris2+}, to match the closest \I2\ profile to our one, we use the \mgii\ and the triplet lines.
EBs are usually found in active regions, where the \mgii\ lines are enhanced. Thus, the most common scenario is to find the wings of the triplet and the \mgii\ enhanced, which we expect to be well sampled by \I2. 
However, we have rare cases, with profiles alike the EBs from 2023/07/27, where the triplet wings are enhanced but the \mgii\ lines are not. These kinds of profiles might not be sampled by \I2\ due to its scarcity. 
Then, the \mgii\ lines dominate over the triplet during the matching.
The closest profile from \I2\ matches the \mgii\ lines but not the triplet, showing no imprint of it in the atmosphere.
A possible fix to this would be to refine the inversions, focusing on the triplet line.
Nonetheless, the statistics shown in Fig.~\ref{fig:ebs_atmos_params_all} demonstrate that this problem occurs for a non-significant number of pixels, since the general trend is very well constrained for the temperature.

\subsection{\Ha\ - triplet relation}

The co-existence of spectral signatures in the \Ha\ and in the IRIS NUV in EBs points to the possibility of a correlation between both lines. 
A strong correlation would imply similar formation conditions.
In \citet{2017ApJ...838..101H}, they correlated the integrated intensity of the triplet with the integration of the \Ha\ line for a strong EB. They obtained very strong correlations, with Pearson correlation coefficients of 0.85 for the triplet-\Ha\ integrations.
Our results, presented in the left panel of Fig.~\ref{fig:correlation_matrix}, are much lower. We obtained correlations with Pearson coefficients between 0.54 and 0.61.
We also computed the correlation using the integrated intensity of the triplet and \Ha\ lines. However, the correlation obtained was even lower.
The causes for the differences can be diverse. Our result comes from 18 EBs from four different observations, while \citet{2017ApJ...838..101H} used one EB. 
A larger amount of data implies more variation, which can lower the linear correlation. 
We assumed the relation between \Ha\ and the triplet to be linear. However, this may not be the case. 
Although both lines showed response to EBs, how these vary with the different thermodynamic parameters of the atmosphere can be different.
If there is a real correlation, it may be non-linear.
Another factor to take into account is the precision of SST-IRIS data alignment, which can induce offsets.

\subsection{EBs detection using IRIS}

A fundamental question of this paper is the possibility of detecting EBs solely using IRIS.
%Grubecka and Tiago paper
Some studies worked with the characterization of \mgii\ compact brightenings, defining them in different ways. They claimed that those profiles could be EBs, but due to their lack of \Ha\ co-observations, it was not possible to fully determine if the events were actually EBs.
Comparing with our EB profiles, we agree that the triplet emission profiles shown in Fig.~9 of \citet{2025A&A...700A.214L}, and 
profiles CB~A, B, C, and D in Fig.~5 from \citet{2016A&A...593A..32G} could correspond to EBs. However, profile CB~E is more similar to the QSEBs profiles from \citet{2024A&A...689A.156B}.
In the same way, we also consider all the profiles shown in Fig.~7 from \citet{2015ApJ...806...14P} to be produced by EBs.
We developed a recipe solely based on the triplet line, with which we recovered 14 of 18 events associated with the \Ha\ EBs, with 1 false positive. 
However, the NUV EBs we recovered are not a perfect match of \Ha\ EBs, but compact and transient events with some temporal and spatial overlap.
This result was expected because although the triplet and the \Ha\ share some similarities (see Fig.~\ref{fig:correlation_matrix}), they are different lines. 
A further study of the full disk mosaics produced by IRIS would be key to understanding EB's contribution to the Sun as a whole.

\section{Concluding remarks}

We have studied EB signatures in the IRIS NUV spectra, combining four different co-temporal and co-spatial SST~\Ha\ and IRIS~NUV observations of active regions.
From the analysis of 18 EBs, we conclude that the triplet line is the main and most reliable IRIS NUV tracer.
This signature manifests as an enhancement of the triplet wings while the core remains in absorption, which is produced by a temperature increase of $\Delta T\sim~1670$~K around $\ltau=-3.8$.
We demonstrated that the spectral behavior of the triplet can also be used to detect \Ha\ EB-associated events.
Furthermore, the \mgii\ lines and the bump between \mgii\ can be used in combination with the triplet as a proxy for the height of formation of the EBs.
By establishing these IRIS NUV spectral proxies, this study highlights the importance of space-based UV observatories such as IRIS, SDO, Solar Orbiter \citep[SolO;][]{2020A&A...642A...1M}, or the upcoming Multi-Slit Solar Explorer \citep[MUSE;][]{2019AGUFMSH33A..07D}, in providing continuous, large-scale diagnostics ---not achievable by ground-based facilities--- key to understanding the energetic processes of the Sun.

\begin{acknowledgements}
% {We would like to thank the anonymous referee for their comments and suggestions.}

This research is supported by the Research Council of Norway, project numbers 358540, % OESE, 325491, % ISSRESS (Luc)
361159, % Reetika NFR Recruitment of Talented Researcher
and through its Centres of Excellence scheme, project number 262622.% RoCS (almost all of us)
This project has also received funding from the European Union's Horizon 2020 research and innovation programme under the Marie Skłodowska-Curie grant agreement Nº 945371;%COMPSCI (Ignasi)
 as well as from the European Research Council through the Synergy Grant number 810218 (``The Whole Sun'', ERC-2018-SyG). %ERC (Daniel)
%
%RJ is supported through the project number 361159. % NFR Recruitment of Talented Researcher
The Swedish 1-m Solar Telescope (SST) is operated on the island of La Palma by the Institute for Solar Physics of Stockholm University in the Spanish Observatorio del Roque de los Muchachos of the Instituto de Astrofísica de Canarias. 
The SST is co-funded by the Swedish Research Council as a national research infrastructure (registration number 4.3-2021-00169).
ASD acknowledges support from NASA contract NNG09FA40C (IRIS). IRIS is a NASA small explorer mission developed and operated by LMSAL, with mission operations executed at NASA Ames Research Center, and major contributions to downlink communications funded by ESA and the Norwegian Space Agency.
We acknowledge the community effort devoted to the development of the following open-source packages that were used in this work: numpy (\url{numpy.org}), matplotlib (\url{matplotlib.org}), scipy (\url{scipy.org}), and astropy (\url{astropy.org}).
This research has made use of NASA's Astrophysics Data System Bibliographic Services.
\end{acknowledgements}

\bibliography{iris_nuv_eb.bib}{}

@article{2021A&A...647A.188D,
	adsurl = {https://ui.adsabs.harvard.edu/abs/2021A&A...647A.188D},
	archiveprefix = {arXiv},
	author = {{D{\'\i}az Baso}, C.~J. and {de la Cruz Rodr{\'\i}guez}, J. and {Leenaarts}, J.},
	doi = {10.1051/0004-6361/202040111},
	eid = {A188},
	eprint = {2012.06229},
	journal = {\aap},
	month = mar,
	pages = {A188},
	primaryclass = {astro-ph.SR},
	title = {{An observationally constrained model of strong magnetic reconnection in the solar chromosphere. Atmospheric stratification and estimates of heating rates}},
	volume = {647},
	year = 2021}

@article{2011ApJ...736...71W,
	adsurl = {https://ui.adsabs.harvard.edu/abs/2011ApJ...736...71W},
	archiveprefix = {arXiv},
	author = {{Watanabe}, Hiroko and {Vissers}, Gregal and {Kitai}, Reizaburo and {Rouppe van der Voort}, Luc and {Rutten}, Robert J.},
	doi = {10.1088/0004-637X/736/1/71},
	eid = {71},
	eprint = {1105.4008},
	journal = {\apj},
	month = jul,
	number = {1},
	pages = {71},
	primaryclass = {astro-ph.SR},
	title = {{Ellerman Bombs at High Resolution. I. Morphological Evidence for Photospheric Reconnection}},
	volume = {736},
	year = 2011}

@article{2013ApJ...774...32V,
	adsurl = {https://ui.adsabs.harvard.edu/abs/2013ApJ...774...32V},
	archiveprefix = {arXiv},
	author = {{Vissers}, Gregal. and {Rouppe van der Voort}, Luc. and {Rutten}, Robert.},
	doi = {10.1088/0004-637X/774/1/32},
	eid = {32},
	eprint = {1307.1547},
	journal = {\apj},
	month = sep,
	number = {1},
	pages = {32},
	primaryclass = {astro-ph.SR},
	title = {{Ellerman Bombs at High Resolution. II. Triggering, Visibility, and Effect on Upper Atmosphere}},
	volume = {774},
	year = 2013}

@article{2002ApJ...575..506G,
	adsurl = {https://ui.adsabs.harvard.edu/abs/2002ApJ...575..506G},
	author = {{Georgoulis}, Manolis. and {Rust}, David. and {Bernasconi}, Pietro. and {Schmieder}, Brigitte},
	doi = {10.1086/341195},
	journal = {\apj},
	month = aug,
	number = {1},
	pages = {506-528},
	title = {{Statistics, Morphology, and Energetics of Ellerman Bombs}},
	volume = {575},
	year = 2002}

@article{2019A&A...626A...4V,
	adsurl = {https://ui.adsabs.harvard.edu/abs/2019A&A...626A...4V},
	archiveprefix = {arXiv},
	author = {{Vissers}, Gregal. and {Rouppe van der Voort}, Luc. and {Rutten}, Robert.},
	doi = {10.1051/0004-6361/201834811},
	eid = {A4},
	eprint = {1901.07975},
	journal = {\aap},
	month = jun,
	pages = {A4},
	primaryclass = {astro-ph.SR},
	title = {{Automating Ellerman bomb detection in ultraviolet continua}},
	volume = {626},
	year = 2019}

@article{2022A&A...664A..72J,
	adsurl = {https://ui.adsabs.harvard.edu/abs/2022A&A...664A..72J},
	archiveprefix = {arXiv},
	author = {{Joshi}, Jayant and {Rouppe van der Voort}, Luc H.~M.},
	doi = {10.1051/0004-6361/202243051},
	eid = {A72},
	eprint = {2203.08172},
	journal = {\aap},
	month = aug,
	pages = {A72},
	primaryclass = {astro-ph.SR},
	title = {{Properties of ubiquitous magnetic reconnection events in the lower solar atmosphere}},
	volume = {664},
	year = 2022}

@INPROCEEDINGS{2013JPhCS.440a2007R,
       author = {{Rutten}, Robert J. and {Vissers}, Gregal J. and {Rouppe van der Voort}, Luc and {S{\"u}tterlin}, Peter and {Vitas}, Nikola},
        title = "{Ellerman bombs: fallacies, fads, usage}",
    booktitle = {J. Phys. Conf. Ser.},
         year = 2013,
       volume = {440},
        month = jun,
    publisher = {IOP},
          eid = {012007},
        pages = {012007},
          doi = {10.1088/1742-6596/440/1/012007},
archivePrefix = {arXiv},
       eprint = {1304.1364},
 primaryClass = {astro-ph.SR},
       adsurl = {https://ui.adsabs.harvard.edu/abs/2013JPhCS.440a2007R}
}

@article{ellerman_solar_1917,
	author = {Ellerman, Ferdinand},
	doi = {10.1086/142366},
	issn = {0004-637X},
	journal = {\apj},
	month = nov,
	note = {ADS Bibcode: 1917ApJ....46..298E},
	pages = {298},
	title = {Solar {Hydrogen} "bombs"},
	url = {https://ui.adsabs.harvard.edu/abs/1917ApJ....46..298E},
	urldate = {2022-10-20},
	volume = {46},
	year = {1917}}

@article{2023A&A...677A..52K,
	adsurl = {https://ui.adsabs.harvard.edu/abs/2023A&A...677A..52K},
	archiveprefix = {arXiv},
	author = {{Krikova}, K. and {Pereira}, T. and {Rouppe van der Voort}, L.},
	doi = {10.1051/0004-6361/202346796},
	eid = {A52},
	eprint = {2307.11131},
	journal = {\aap},
	month = sep,
	pages = {A52},
	primaryclass = {astro-ph.SR},
	title = {{Formation of H{\ensuremath{\varepsilon}} in the solar atmosphere}},
	volume = {677},
	year = 2023}

@article{2024A&A...683A.190R,
	adsurl = {https://ui.adsabs.harvard.edu/abs/2024A&A...683A.190R},
	archiveprefix = {arXiv},
	author = {{Rouppe van der Voort}, Luc H.~M. and {Joshi}, Jayant and {Krikova}, Kilian},
	doi = {10.1051/0004-6361/202348976},
	eid = {A190},
	eprint = {2401.12077},
	journal = {\aap},
	month = mar,
	pages = {A190},
	primaryclass = {astro-ph.SR},
	title = {{Observations of magnetic reconnection in the deep solar atmosphere in the H{\ensuremath{\varepsilon}} line}},
	volume = {683},
	year = 2024}

@ARTICLE{2006SoPh..235...75S,
       author = {{Socas-Navarro}, Hector and {Pillet}, Valent{\'\i}n Mart{\'\i}nez and {Elmore}, David and {Pietarila}, Anna and {Lites}, Bruce W. and {Sainz}, Rafael Manso},
        title = "{Spectro-Polarimetric Observations and Non-Lte Modeling of Ellerman Bombs}",
      journal = {\solphys},
         year = 2006,
        month = may,
       volume = {235},
       number = {1-2},
        pages = {75-86},
          doi = {10.1007/s11207-006-0049-x},
archivePrefix = {arXiv},
       eprint = {astro-ph/0508667},
 primaryClass = {astro-ph},
       adsurl = {https://ui.adsabs.harvard.edu/abs/2006SoPh..235...75S}
}

@article{2007A&A...473..279P,
	adsurl = {https://ui.adsabs.harvard.edu/abs/2007A&A...473..279P},
	author = {{Pariat}, E. and {Schmieder}, B. and {Berlicki}, A. and {Deng}, Y. and {Mein}, N. and {L{\'o}pez Ariste}, A. and {Wang}, S.},
	doi = {10.1051/0004-6361:20067011},
	journal = {\aap},
	month = oct,
	number = {1},
	pages = {279-289},
	title = {{Spectrophotometric analysis of Ellerman bombs in the Ca II, H{\ensuremath{\alpha}}, and UV range}},
	volume = {473},
	year = 2007}

@article{matsumoto_CaH,
	author = {Matsumoto, Takuma and Kitai, R. and Shibata, Kazunari and Nagata, Shin'ichi and Otsuji, Kenichi and Nakamura, Tahei and Watanabe, Hiroko and Tsuneta, Saku and Suematsu, Yoshinori and Ichimoto, Kiyoshi and Shimizu, Toshifumi and Katsukawa, Yumi and Tarbell, Ted and Lites, Bruce and Shine, Richard and Title, Alan},
	doi = {10.1093/pasj/60.3.577},
	journal = {\pasj},
	month = {05},
	pages = {577},
	title = {Cooperative Observation of Ellerman Bombs between the Solar Optical Telescope aboard Hinode and Hida/Domeless Solar Telescope},
	volume = {60},
	year = {2008}}

@article{2019A&A...627A.101V,
	adsurl = {https://ui.adsabs.harvard.edu/abs/2019A&A...627A.101V},
	archiveprefix = {arXiv},
	author = {{Vissers}, G. and {de la Cruz Rodr{\'\i}guez}, J. and {Libbrecht}, T. and {Rouppe van der Voort}, L. and {Scharmer}, G. and {Carlsson}, M.},
	doi = {10.1051/0004-6361/201833560},
	eid = {A101},
	eprint = {1905.02035},
	journal = {\aap},
	month = jul,
	pages = {A101},
	primaryclass = {astro-ph.SR},
	title = {{Dissecting bombs and bursts: non-LTE inversions of low-atmosphere reconnection in SST and IRIS observations}},
	volume = {627},
	year = 2019}

@article{2012SoPh..275...17L,
	adsurl = {https://ui.adsabs.harvard.edu/abs/2012SoPh..275...17L},
	author = {{Lemen}, James R. and {Title}, Alan M. and {Akin}, David J. and {Boerner}, Paul F. and {Chou}, Catherine and {Drake}, Jerry F. and {Duncan}, Dexter W. and {Edwards}, Christopher G. and {Friedlaender}, Frank M. and {Heyman}, Gary F. and {Hurlburt}, Neal E. and {Katz}, Noah L. and {Kushner}, Gary D. and {Levay}, Michael and {Lindgren}, Russell W. and {Mathur}, Dnyanesh P. and {McFeaters}, Edward L. and {Mitchell}, Sarah and {Rehse}, Roger A. and {Schrijver}, Carolus J. and {Springer}, Larry A. and {Stern}, Robert A. and {Tarbell}, Theodore D. and {Wuelser}, Jean-Pierre and {Wolfson}, C. Jacob and {Yanari}, Carl and {Bookbinder}, Jay A. and {Cheimets}, Peter N. and {Caldwell}, David and {Deluca}, Edward E. and {Gates}, Richard and {Golub}, Leon and {Park}, Sang and {Podgorski}, William A. and {Bush}, Rock I. and {Scherrer}, Philip H. and {Gummin}, Mark A. and {Smith}, Peter and {Auker}, Gary and {Jerram}, Paul and {Pool}, Peter and {Soufli}, Regina and {Windt}, David L. and {Beardsley}, Sarah and {Clapp}, Matthew and {Lang}, James and {Waltham}, Nicholas},
	doi = {10.1007/s11207-011-9776-8},
	journal = {\solphys},
	month = jan,
	number = {1-2},
	pages = {17-40},
	title = {{The Atmospheric Imaging Assembly (AIA) on the Solar Dynamics Observatory (SDO)}},
	volume = {275},
	year = 2012}

@INPROCEEDINGS{2003SPIE.4853..341S,
   author = {{Scharmer}, G.~B. and {Bjelksj{\"o}}, K. and {Korhonen}, T.~K. and 
	{Lindberg}, B. and {Petterson}, B.},
    title = "{The 1-meter Swedish solar telescope}",
booktitle = {Innovative Telescopes and Instrumentation for Solar Astrophysics},
     year = 2003,
   series = {\procspie},
   volume = 4853,
   editor = {{Keil}, S.~L. and {Avakyan}, S.~V.},
    month = feb,
    pages = {341-350},
      doi = {10.1117/12.460377},
   adsurl = {http://adsabs.harvard.edu/abs/2003SPIE.4853..341S}
}

@article{2020A&A...642A...1M,
	adsurl = {https://ui.adsabs.harvard.edu/abs/2020A&A...642A...1M},
	archiveprefix = {arXiv},
	author = {{M{\"u}ller}, D. and {St. Cyr}, O.~C. and {Zouganelis}, I. and {Gilbert}, H.~R. and {Marsden}, R. and {Nieves-Chinchilla}, T. and {Antonucci}, E. and {Auch{\`e}re}, F. and {Berghmans}, D. and {Horbury}, T.~S. and {Howard}, R.~A. and {Krucker}, S. and {Maksimovic}, M. and {Owen}, C.~J. and {Rochus}, P. and {Rodriguez-Pacheco}, J. and {Romoli}, M. and {Solanki}, S.~K. and {Bruno}, R. and {Carlsson}, M. and {Fludra}, A. and {Harra}, L. and {Hassler}, D.~M. and {Livi}, S. and {Louarn}, P. and {Peter}, H. and {Sch{\"u}hle}, U. and {Teriaca}, L. and {del Toro Iniesta}, J.~C. and {Wimmer-Schweingruber}, R.~F. and {Marsch}, E. and {Velli}, M. and {De Groof}, A. and {Walsh}, A. and {Williams}, D.},
	doi = {10.1051/0004-6361/202038467},
	eid = {A1},
	eprint = {2009.00861},
	journal = {\aap},
	month = oct,
	pages = {A1},
	primaryclass = {astro-ph.SR},
	title = {{The Solar Orbiter mission. Science overview}},
	volume = {642},
	year = 2020}

@article{2021ApJ...921...50H,
	adsurl = {https://ui.adsabs.harvard.edu/abs/2021ApJ...921...50H},
	archiveprefix = {arXiv},
	author = {{Hong}, Jie and {Li}, Ying and {Ding}, M.~D. and {Hao}, Qi},
	doi = {10.3847/1538-4357/ac1ba0},
	eid = {50},
	eprint = {2108.02699},
	journal = {\apj},
	month = nov,
	number = {1},
	pages = {50},
	primaryclass = {astro-ph.SR},
	title = {{Revisiting the Spectral Features of Ellerman Bombs and UV Bursts. I. Radiative Hydrodynamic Simulations}},
	volume = {921},
	year = 2021}

@article{2019A&A...626A..33H,
	adsurl = {https://ui.adsabs.harvard.edu/abs/2019A&A...626A..33H},
	archiveprefix = {arXiv},
	author = {{Hansteen}, V. and {Ortiz}, A. and {Archontis}, V. and {Carlsson}, M. and {Pereira}, T. and {Bj{\o}rgen}, J.},
	doi = {10.1051/0004-6361/201935376},
	eid = {A33},
	eprint = {1904.11524},
	journal = {\aap},
	month = jun,
	pages = {A33},
	primaryclass = {astro-ph.SR},
	title = {{Ellerman bombs and UV bursts: transient events in chromospheric current sheets}},
	volume = {626},
	year = 2019}

@article{kleint2022occurrence,
	author = {Kleint, Lucia and Panos, Brandon},
	journal = {Astronomy \& Astrophysics},
	pages = {A132},
	publisher = {EDP Sciences},
	title = {Occurrence and statistics of IRIS bursts},
	volume = {657},
	year = {2022}}

@article{2008ApJ...689L..69S,
	adsurl = {https://ui.adsabs.harvard.edu/abs/2008ApJ...689L..69S},
	archiveprefix = {arXiv},
	author = {{Scharmer}, G.~B. and {Narayan}, G. and {Hillberg}, T. and {de la Cruz Rodriguez}, J. and {L{\"o}fdahl}, M.~G. and {Kiselman}, D. and {S{\"u}tterlin}, P. and {van Noort}, M. and {Lagg}, A.},
	doi = {10.1086/595744},
	eprint = {0806.1638},
	journal = {\apjl},
	month = dec,
	number = {1},
	pages = {L69},
	primaryclass = {astro-ph},
	title = {{CRISP Spectropolarimetric Imaging of Penumbral Fine Structure}},
	volume = {689},
	year = 2008}

@article{2015A&A...573A..40D,
	adsurl = {https://ui.adsabs.harvard.edu/abs/2015A&A...573A..40D},
	archiveprefix = {arXiv},
	author = {{de la Cruz Rodr{\'\i}guez}, J. and {L{\"o}fdahl}, M.~G. and {S{\"u}tterlin}, P. and {Hillberg}, T. and {Rouppe van der Voort}, L.},
	doi = {10.1051/0004-6361/201424319},
	eid = {A40},
	eprint = {1406.0202},
	journal = {\aap},
	month = jan,
	pages = {A40},
	primaryclass = {astro-ph.SR},
	title = {{CRISPRED: A data pipeline for the CRISP imaging spectropolarimeter}},
	volume = {573},
	year = 2015}

@article{2021A&A...653A..68L,
	adsurl = {https://ui.adsabs.harvard.edu/abs/2021A&A...653A..68L},
	archiveprefix = {arXiv},
	author = {{L{\"o}fdahl}, Mats G. and {Hillberg}, Tomas and {de la Cruz Rodr{\'\i}guez}, Jaime and {Vissers}, Gregal and {Andriienko}, Oleksii and {Scharmer}, G{\"o}ran B. and {Haugan}, Stein V.~H. and {Fredvik}, Terje},
	doi = {10.1051/0004-6361/202141326},
	eid = {A68},
	eprint = {1804.03030},
	journal = {\aap},
	month = sep,
	pages = {A68},
	primaryclass = {astro-ph.IM},
	title = {{SSTRED: Data- and metadata-processing pipeline for CHROMIS and CRISP}},
	volume = {653},
	year = 2021}

@article{2005SoPh..228..191V,
	adsurl = {https://ui.adsabs.harvard.edu/abs/2005SoPh..228..191V},
	author = {{Van Noort}, Michiel and {Rouppe van der Voort}, Luc and {L{\"o}fdahl}, Mats.},
	doi = {10.1007/s11207-005-5782-z},
	journal = {\solphys},
	month = may,
	number = {1-2},
	pages = {191-215},
	title = {{Solar Image Restoration By Use Of Multi-frame Blind De-convolution With Multiple Objects And Phase Diversity}},
	volume = {228},
	year = 2005}

@article{2015ApJ...812...11V,
	adsurl = {https://ui.adsabs.harvard.edu/abs/2015ApJ...812...11V},
	archiveprefix = {arXiv},
	author = {{Vissers}, G. and {Rouppe van der Voort}, L. and {Rutten}, R. and {Carlsson}, M. and {De Pontieu}, B.},
	doi = {10.1088/0004-637X/812/1/11},
	eid = {11},
	eprint = {1507.00435},
	journal = {\apj},
	month = oct,
	number = {1},
	pages = {11},
	primaryclass = {astro-ph.SR},
	title = {{Ellerman Bombs at High Resolution. III. Simultaneous Observations with IRIS and SST}},
	volume = {812},
	year = 2015}

@inproceedings{2019AGUFMSH33A..07D,
	adsurl = {https://ui.adsabs.harvard.edu/abs/2019AGUFMSH33A..07D},
	author = {{De Pontieu}, B. and {Lemen}, J.~R. and {Cheung}, C.~M.~M.},
	booktitle = {AGU Fall Meeting Abstracts},
	eid = {SH33A-07},
	month = dec,
	pages = {SH33A-07},
	title = {{MUSE: the Multi-Slit Solar Explorer}},
	volume = {2019},
	year = 2019}

@article{2013ApJ...778..143P,
	adsurl = {https://ui.adsabs.harvard.edu/abs/2013ApJ...778..143P},
	archiveprefix = {arXiv},
	author = {{Pereira}, T. and {Leenaarts}, J. and {De Pontieu}, B. and {Carlsson}, M. and {Uitenbroek}, H.},
	doi = {10.1088/0004-637X/778/2/143},
	eid = {143},
	eprint = {1310.1926},
	journal = {\apj},
	month = dec,
	number = {2},
	pages = {143},
	primaryclass = {astro-ph.SR},
	title = {{The Formation of IRIS Diagnostics. III. Near-ultraviolet Spectra and Images}},
	volume = {778},
	year = 2013}

@article{2015ApJ...806...14P,
	adsurl = {https://ui.adsabs.harvard.edu/abs/2015ApJ...806...14P},
	archiveprefix = {arXiv},
	author = {{Pereira}, Tiago M.~D. and {Carlsson}, Mats and {De Pontieu}, Bart and {Hansteen}, Viggo},
	doi = {10.1088/0004-637X/806/1/14},
	eid = {14},
	eprint = {1504.01733},
	journal = {\apj},
	month = jun,
	number = {1},
	pages = {14},
	primaryclass = {astro-ph.SR},
	title = {{The Formation of IRIS Diagnostics. IV. The Mg II Triplet Lines as a New Diagnostic for Lower Chromospheric Heating}},
	volume = {806},
	year = 2015}

@article{ortiz_ellerman_2020,
	author = {Ortiz, Ada and Hansteen, Viggo. and N{\'o}brega-Siverio, Daniel and Rouppe van der Voort, Luc},
	copyright = {{\copyright} ESO 2020},
	doi = {10.1051/0004-6361/201936574},
	issn = {0004-6361, 1432-0746},
	journal = {\aap},
	language = {en},
	month = jan,
	note = {Publisher: EDP Sciences},
	pages = {A58},
	shorttitle = {Ellerman bombs and {UV} bursts},
	title = {Ellerman bombs and {UV} bursts: reconnection at different atmospheric layers},
	url = {https://www.aanda.org/articles/aa/abs/2020/01/aa36574-19/aa36574-19.html},
	urldate = {2023-05-23},
	volume = {633},
	year = {2020}}

@article{2013ApJ...772...90L,
	adsurl = {https://ui.adsabs.harvard.edu/abs/2013ApJ...772...90L},
	archiveprefix = {arXiv},
	author = {{Leenaarts}, J. and {Pereira}, T. and {Carlsson}, M. and {Uitenbroek}, H. and {De Pontieu}, B.},
	doi = {10.1088/0004-637X/772/2/90},
	eid = {90},
	eprint = {1306.0671},
	journal = {\apj},
	month = aug,
	number = {2},
	pages = {90},
	primaryclass = {astro-ph.SR},
	title = {{The Formation of IRIS Diagnostics. II. The Formation of the Mg II h\&k Lines in the Solar Atmosphere}},
	volume = {772},
	year = 2013}

@article{2020ApJ...891...17P,
	adsurl = {https://ui.adsabs.harvard.edu/abs/2020ApJ...891...17P},
	archiveprefix = {arXiv},
	author = {{Panos}, Brandon and {Kleint}, Lucia},
	doi = {10.3847/1538-4357/ab700b},
	eid = {17},
	eprint = {1911.12621},
	journal = {\apj},
	month = mar,
	number = {1},
	pages = {17},
	primaryclass = {astro-ph.SR},
	title = {{Real-time Flare Prediction Based on Distinctions between Flaring and Non-flaring Active Region Spectra}},
	volume = {891},
	year = 2020}

@ARTICLE{2024A&A...686A.218N,
       author = {{N{\'o}brega-Siverio}, D. and {Cabello}, I. and {Bose}, S. and {Rouppe van der Voort}, L.~H.~M. and {Joshi}, R. and {Froment}, C. and {Henriques}, V.~M.~J.},
        title = "{Small-scale magnetic flux emergence preceding a chain of energetic solar atmospheric events}",
      journal = {\aap},
         year = 2024,
        month = jun,
       volume = {686},
          eid = {A218},
        pages = {A218},
          doi = {10.1051/0004-6361/202348894},
archivePrefix = {arXiv},
       eprint = {2403.11652},
 primaryClass = {astro-ph.SR},
       adsurl = {https://ui.adsabs.harvard.edu/abs/2024A&A...686A.218N}
}

@article{2024A&A...689A..72Z,
	adsurl = {https://ui.adsabs.harvard.edu/abs/2024A&A...689A..72Z},
	author = {{Zbinden}, Jonas and {Kleint}, Lucia and {Panos}, Brandon},
	doi = {10.1051/0004-6361/202347824},
	eid = {A72},
	journal = {\aap},
	month = sep,
	pages = {A72},
	title = {{Investigating and comparing the IRIS spectral lines Mg II, Si IV, and C II for flare precursor diagnostics}},
	volume = {689},
	year = 2024}

@article{2021A&A...652A..78S,
	adsurl = {https://ui.adsabs.harvard.edu/abs/2021A&A...652A..78S},
	author = {{Socas-Navarro}, H. and {Asensio Ramos}, A.},
	doi = {10.1051/0004-6361/202140424},
	eid = {A78},
	journal = {\aap},
	month = aug,
	pages = {A78},
	title = {{Exploring the Sun's upper atmosphere with neural networks: Reversed patterns and the hot wall effect}},
	volume = {652},
	year = 2021}

@article{2025A&A...699A..54S,
	adsurl = {https://ui.adsabs.harvard.edu/abs/2025A&A...699A..54S},
	archiveprefix = {arXiv},
	author = {{Soler Poquet}, I. and {D{\'\i}az Baso}, C. and {Rouppe van der Voort}, L. and {Vissers}, G.},
	doi = {10.1051/0004-6361/202453524},
	eid = {A54},
	eprint = {2505.03023},
	journal = {\aap},
	month = jul,
	pages = {A54},
	primaryclass = {astro-ph.SR},
	title = {{Automatic detection of Ellerman bombs using deep learning}},
	volume = {699},
	year = 2025}

@article{2024ApJS..271...24S,
	adsurl = {https://ui.adsabs.harvard.edu/abs/2024ApJS..271...24S},
	archiveprefix = {arXiv},
	author = {{Sainz Dalda}, Alberto and {Agrawal}, Aaryan and {De Pontieu}, Bart and {Go{\v{s}}i{\'c}}, Milan},
	doi = {10.3847/1538-4365/ad1e55},
	eid = {24},
	eprint = {2211.09103},
	journal = {\apjs},
	month = mar,
	number = {1},
	pages = {24},
	primaryclass = {astro-ph.SR},
	title = {{IRIS$^{2+}$: A Comprehensive Database of Stratified Thermodynamic Models in the Low Solar Atmosphere}},
	volume = {271},
	year = 2024}

@article{2017SSRv..210..109D,
	adsurl = {https://ui.adsabs.harvard.edu/abs/2017SSRv..210..109D},
	archiveprefix = {arXiv},
	author = {{de la Cruz Rodr{\'\i}guez}, J. and {van Noort}, M.},
	doi = {10.1007/s11214-016-0294-8},
	eprint = {1609.08324},
	journal = {\ssr},
	month = sep,
	number = {1-4},
	pages = {109-143},
	primaryclass = {astro-ph.SR},
	title = {{Radiative Diagnostics in the Solar Photosphere and Chromosphere}},
	volume = {210},
	year = 2017}

@ARTICLE{2016LRSP...13....4D,
       author = {{del Toro Iniesta}, Jose Carlos and {Ruiz Cobo}, Basilio},
        title = "{Inversion of the radiative transfer equation for polarized light}",
      journal = {Living Reviews in Solar Physics},
         year = 2016,
        month = dec,
       volume = {13},
       number = {1},
          eid = {4},
        pages = {4},
          doi = {10.1007/s41116-016-0005-2},
archivePrefix = {arXiv},
       eprint = {1610.10039},
 primaryClass = {astro-ph.SR},
       adsurl = {https://ui.adsabs.harvard.edu/abs/2016LRSP...13....4D}
}

@ARTICLE{stic,
       author = {{de la Cruz Rodr{\'\i}guez}, J. and {Leenaarts}, J. and {Danilovic}, S. and {Uitenbroek}, H.},
        title = "{STiC: A multiatom non-LTE PRD inversion code for full-Stokes solar observations}",
      journal = {\aap},
         year = 2019,
        month = mar,
       volume = {623},
          eid = {A74},
        pages = {A74},
          doi = {10.1051/0004-6361/201834464},
archivePrefix = {arXiv},
       eprint = {1810.08441},
 primaryClass = {astro-ph.SR},
       adsurl = {https://ui.adsabs.harvard.edu/abs/2019A&A...623A..74D}
}

@article{2016ApJ...824...96T,
	adsurl = {https://ui.adsabs.harvard.edu/abs/2016ApJ...824...96T},
	archiveprefix = {arXiv},
	author = {{Tian}, Hui and {Xu}, Zhi and {He}, Jiansen and {Madsen}, Chad},
	doi = {10.3847/0004-637X/824/2/96},
	eid = {96},
	eprint = {1604.05423},
	journal = {\apj},
	month = jun,
	number = {2},
	pages = {96},
	primaryclass = {astro-ph.SR},
	title = {{Are IRIS Bombs Connected to Ellerman Bombs?}},
	volume = {824},
	year = 2016}

@article{2016A&A...592A.100R,
	adsurl = {https://ui.adsabs.harvard.edu/abs/2016A&A...592A.100R},
	archiveprefix = {arXiv},
	author = {{Rouppe van der Voort}, Luc. and {Rutten}, Robert. and {Vissers}, Gregal.},
	doi = {10.1051/0004-6361/201628889},
	eid = {A100},
	eprint = {1606.03675},
	journal = {\aap},
	month = aug,
	pages = {A100},
	primaryclass = {astro-ph.SR},
	title = {{Reconnection brightenings in the quiet solar photosphere}},
	volume = {592},
	year = 2016}

@article{2024A&A...689A.156B,
	adsurl = {https://ui.adsabs.harvard.edu/abs/2024A&A...689A.156B},
	archiveprefix = {arXiv},
	author = {{Bhatnagar}, Aditi and {Rouppe van der Voort}, Luc and {Joshi}, Jayant},
	doi = {10.1051/0004-6361/202450070},
	eid = {A156},
	eprint = {2406.09585},
	journal = {\aap},
	month = sep,
	pages = {A156},
	primaryclass = {astro-ph.SR},
	title = {{Transition region response to quiet-Sun Ellerman bombs}},
	volume = {689},
	year = 2024}

@article{2017ApJ...845...16N,
	adsurl = {https://ui.adsabs.harvard.edu/abs/2017ApJ...845...16N},
	archiveprefix = {arXiv},
	author = {{Nelson}, C.~J. and {Freij}, N. and {Reid}, A. and {Oliver}, R. and {Mathioudakis}, M. and {Erd{\'e}lyi}, R.},
	doi = {10.3847/1538-4357/aa7e7a},
	eid = {16},
	eprint = {1707.05080},
	journal = {\apj},
	month = aug,
	number = {1},
	pages = {16},
	primaryclass = {astro-ph.SR},
	title = {{IRIS Burst Spectra Co-spatial to a Quiet-Sun Ellerman-like Brightening}},
	volume = {845},
	year = 2017}

@article{2017ApJ...838..101H,
	adsurl = {https://ui.adsabs.harvard.edu/abs/2017ApJ...838..101H},
	archiveprefix = {arXiv},
	author = {{Hong}, Jie and {Ding}, M.~D. and {Cao}, Wenda},
	doi = {10.3847/1538-4357/aa671e},
	eid = {101},
	eprint = {1703.04268},
	journal = {\apj},
	month = apr,
	number = {2},
	pages = {101},
	primaryclass = {astro-ph.SR},
	title = {{Multi-wavelength Spectral Analysis of Ellerman Bombs Observed by FISS and IRIS}},
	volume = {838},
	year = 2017}

@article{2019ApJ...875L..30C,
	adsurl = {https://ui.adsabs.harvard.edu/abs/2019ApJ...875L..30C},
	archiveprefix = {arXiv},
	author = {{Chen}, Yajie and {Tian}, Hui and {Peter}, Hardi and {Samanta}, Tanmoy and {Yurchyshyn}, Vasyl and {Wang}, Haimin and {Cao}, Wenda and {Wang}, Linghua and {He}, Jiansen},
	doi = {10.3847/2041-8213/ab18a4},
	eid = {L30},
	eprint = {1903.01981},
	journal = {\apjl},
	month = apr,
	number = {2},
	pages = {L30},
	primaryclass = {astro-ph.SR},
	title = {{Flame-like Ellerman Bombs and Their Connection to Solar Ultraviolet Bursts}},
	volume = {875},
	year = 2019}

@article{2017ApJ...839...22H,
	adsurl = {https://ui.adsabs.harvard.edu/abs/2017ApJ...839...22H},
	archiveprefix = {arXiv},
	author = {{Hansteen}, V.~H. and {Archontis}, V. and {Pereira}, T.~M.~D. and {Carlsson}, M. and {Rouppe van der Voort}, L. and {Leenaarts}, J.},
	doi = {10.3847/1538-4357/aa6844},
	eid = {22},
	eprint = {1704.02872},
	journal = {\apj},
	month = apr,
	number = {1},
	pages = {22},
	primaryclass = {astro-ph.SR},
	title = {{Bombs and Flares at the Surface and Lower Atmosphere of the Sun}},
	volume = {839},
	year = 2017}

@article{2014ApJ...792...13H,
	adsurl = {https://ui.adsabs.harvard.edu/abs/2014ApJ...792...13H},
	archiveprefix = {arXiv},
	author = {{Hong}, Jie and {Ding}, M.~D. and {Li}, Ying and {Fang}, Cheng and {Cao}, Wenda},
	doi = {10.1088/0004-637X/792/1/13},
	eid = {13},
	eprint = {1407.3048},
	journal = {\apj},
	month = sep,
	number = {1},
	pages = {13},
	primaryclass = {astro-ph.SR},
	title = {{Spectral Observations of Ellerman Bombs and Fitting with a Two-cloud Model}},
	volume = {792},
	year = 2014}

@article{2016A&A...593A..32G,
	adsurl = {https://ui.adsabs.harvard.edu/abs/2016A&A...593A..32G},
	author = {{Grubecka Litwicka}, M. and {Schmieder}, B. and {Berlicki}, A. and {Heinzel}, P. and {Dalmasse}, K. and {Mein}, P.},
	doi = {10.1051/0004-6361/201527358},
	eid = {A32},
	journal = {\aap},
	month = sep,
	pages = {A32},
	title = {{Height formation of bright points observed by IRIS in Mg II line wings during flux emergence}},
	volume = {593},
	year = 2016}

@article{2020A&A...641A.146R,
	adsurl = {https://ui.adsabs.harvard.edu/abs/2020A&A...641A.146R},
	archiveprefix = {arXiv},
	author = {{Rouppe van der Voort}, L. and {De Pontieu}, B. and {Carlsson}, M. and {de la Cruz Rodr{\'\i}guez}, J. and {Bose}, S. and {Chintzoglou}, G. and {Drews}, A. and {Froment}, C. and {Go{\v{s}}i{\'c}}, M. and {Graham}, D.~R. and {Hansteen}, V.~H. and {Henriques}, V.~M.~J. and {Jafarzadeh}, S. and {Joshi}, J. and {Kleint}, L. and {Kohutova}, P. and {Leifsen}, T. and {Mart{\'\i}nez-Sykora}, J. and {N{\'o}brega-Siverio}, D. and {Ortiz}, A. and {Pereira}, T.~M.~D. and {Popovas}, A. and {Quintero Noda}, C. and {Sainz Dalda}, A. and {Scharmer}, G.~B. and {Schmit}, D. and {Scullion}, E. and {Skogsrud}, H. and {Szydlarski}, M. and {Timmons}, R. and {Vissers}, G.~J.~M. and {Woods}, M.~M. and {Zacharias}, P.},
	doi = {10.1051/0004-6361/202038732},
	eid = {A146},
	eprint = {2005.14175},
	journal = {\aap},
	month = sep,
	pages = {A146},
	primaryclass = {astro-ph.SR},
	title = {{High-resolution observations of the solar photosphere, chromosphere, and transition region. A database of coordinated IRIS and SST observations}},
	volume = {641},
	year = 2020}

@article{2014SoPh..289.2733D,
	adsurl = {https://ui.adsabs.harvard.edu/abs/2014SoPh..289.2733D},
	archiveprefix = {arXiv},
	author = {{De Pontieu}, B. and {Title}, A.~M. and {Lemen}, J.~R. and {Kushner}, G.~D. and {Akin}, D.~J. and {Allard}, B. and {Berger}, T. and {Boerner}, P. and {Cheung}, M. and {Chou}, C. and {Drake}, J.~F. and {Duncan}, D.~W. and {Freeland}, S. and {Heyman}, G.~F. and {Hoffman}, C. and {Hurlburt}, N.~E. and {Lindgren}, R.~W. and {Mathur}, D. and {Rehse}, R. and {Sabolish}, D. and {Seguin}, R. and {Schrijver}, C.~J. and {Tarbell}, T.~D. and {W{\"u}lser}, J.-P. and {Wolfson}, C.~J. and {Yanari}, C. and {Mudge}, J. and {Nguyen-Phuc}, N. and {Timmons}, R. and {van Bezooijen}, R. and {Weingrod}, I. and {Brookner}, R. and {Butcher}, G. and {Dougherty}, B. and {Eder}, J. and {Knagenhjelm}, V. and {Larsen}, S. and {Mansir}, D. and {Phan}, L. and {Boyle}, P. and {Cheimets}, P.~N. and {DeLuca}, E.~E. and {Golub}, L. and {Gates}, R. and {Hertz}, E. and {McKillop}, S. and {Park}, S. and {Perry}, T. and {Podgorski}, W.~A. and {Reeves}, K. and {Saar}, S. and {Testa}, P. and {Tian}, H. and {Weber}, M. and {Dunn}, C. and {Eccles}, S. and {Jaeggli}, S.~A. and {Kankelborg}, C.~C. and {Mashburn}, K. and {Pust}, N. and {Springer}, L. and {Carvalho}, R. and {Kleint}, L. and {Marmie}, J. and {Mazmanian}, E. and {Pereira}, T.~M.~D. and {Sawyer}, S. and {Strong}, J. and {Worden}, S.~P. and {Carlsson}, M. and {Hansteen}, V.~H. and {Leenaarts}, J. and {Wiesmann}, M. and {Aloise}, J. and {Chu}, K.-C. and {Bush}, R.~I. and {Scherrer}, P.~H. and {Brekke}, P. and {Martinez-Sykora}, J. and {Lites}, B.~W. and {McIntosh}, S.~W. and {Uitenbroek}, H. and {Okamoto}, T.~J. and {Gummin}, M.~A. and {Auker}, G. and {Jerram}, P. and {Pool}, P. and {Waltham}, N.},
	doi = {10.1007/s11207-014-0485-y},
	eprint = {1401.2491},
	journal = {\solphys},
	month = jul,
	number = {7},
	pages = {2733-2779},
	primaryclass = {astro-ph.SR},
	title = {{The Interface Region Imaging Spectrograph (IRIS)}},
	volume = {289},
	year = 2014}

@article{1974ApJ...192..769M,
	adsurl = {https://ui.adsabs.harvard.edu/abs/1974ApJ...192..769M},
	author = {{Milkey}, R.~W. and {Mihalas}, D.},
	doi = {10.1086/153115},
	journal = {\apj},
	month = sep,
	pages = {769-776},
	title = {{Resonance line transfer with partial redistribution: II. The solar Mg II lines.}},
	volume = {192},
	year = 1974}

@article{2026ApJ...997..229S,
	adsurl = {https://ui.adsabs.harvard.edu/abs/2026ApJ...997..229S},
	archiveprefix = {arXiv},
	author = {{Sainz Dalda}, Alberto and {de la Cruz Rodr{\'\i}guez}, Jaime and {Hansteen}, Viggo and {De Pontieu}, Bart and {Go{\v{s}}i{\'c}}, Milan},
	doi = {10.3847/1538-4357/ae274c},
	eid = {229},
	eprint = {2601.09005},
	journal = {\apj},
	month = feb,
	number = {2},
	pages = {229},
	primaryclass = {astro-ph.SR},
	title = {{The IRIS$^{2+}$ Inversion Tool: Recovering the Radiative Losses and the Thermodynamics in the Lower Solar Atmosphere}},
	volume = {997},
	year = 2026}

@ARTICLE{2024A&A...685A..32S,
       author = {{Scharmer}, G.~B. and {Sliepen}, G. and {Sinquin}, J.-C. and {L{\"o}fdahl}, M.~G. and {Lindberg}, B. and {S{\"u}tterlin}, P.},
        title = "{The 85-electrode adaptive optics system of the Swedish 1-m Solar Telescope}",
      journal = {\aap},
         year = 2024,
        month = may,
       volume = {685},
          eid = {A32},
        pages = {A32},
          doi = {10.1051/0004-6361/201936005},
archivePrefix = {arXiv},
       eprint = {2311.13690},
 primaryClass = {astro-ph.IM},
       adsurl = {https://ui.adsabs.harvard.edu/abs/2024A&A...685A..32S}
}

@ARTICLE{2025A&A...700A.214L,
       author = {{Litwicka}, M. and {Berlicki}, A. and {Schmieder}, B.},
        title = "{Statistical analysis of compact brightenings in IRIS Mg II h and k lines}",
      journal = {\aap},
         year = 2025,
        month = aug,
       volume = {700},
          eid = {A214},
        pages = {A214},
          doi = {10.1051/0004-6361/202453134},
       adsurl = {https://ui.adsabs.harvard.edu/abs/2025A&A...700A.214L}
}
\bibliographystyle{aa}

\begin{appendix}
    \onecolumn

\section{Observation details}
\label{appendix:observations}

\begin{figure*}[h!]
\includegraphics[width=1\linewidth]{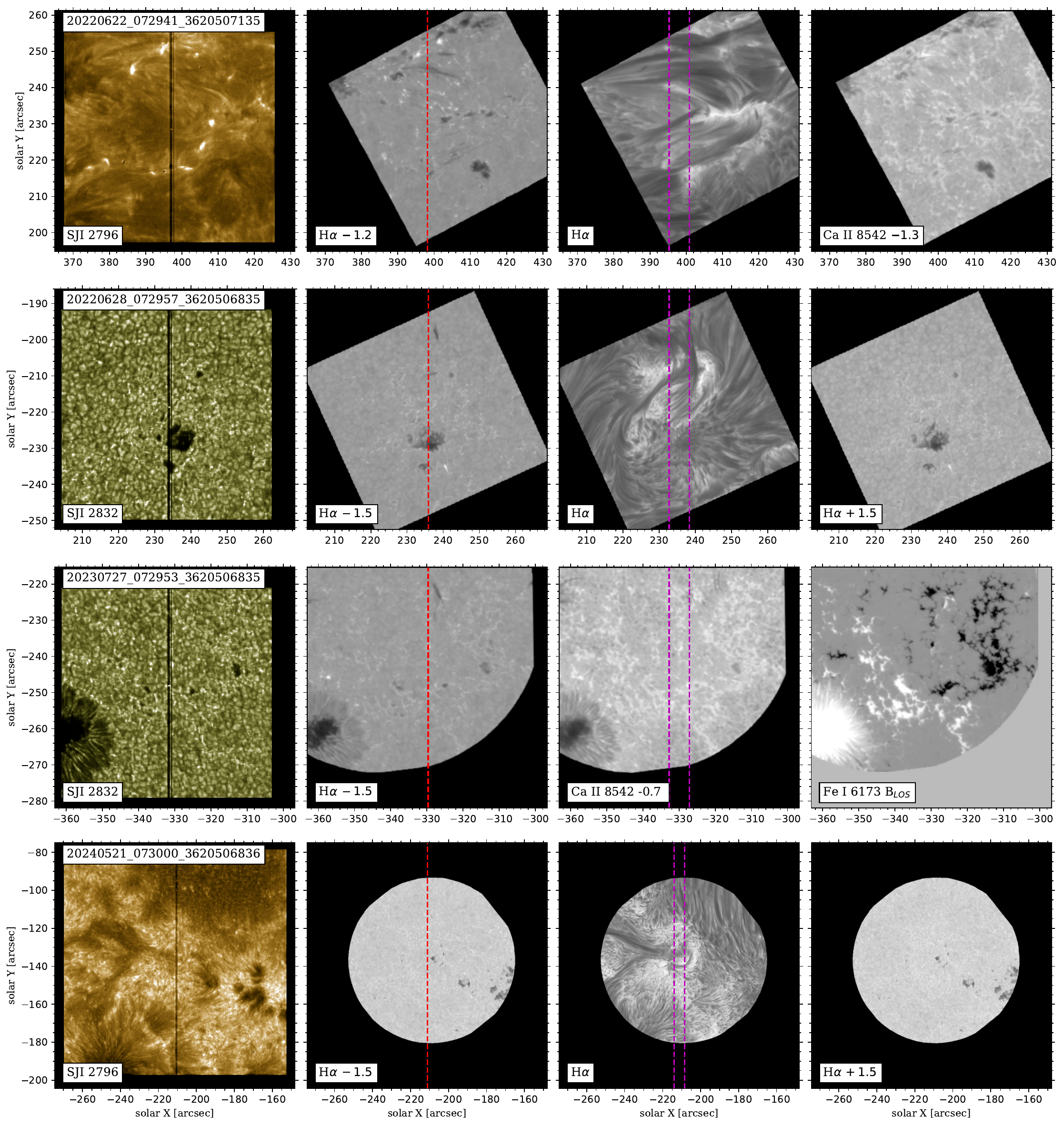}
\caption{Sample images from the four different datasets used. Each row shows four different diagnostics for the same observation. The first column shows an IRIS SJI image, with the corresponding OBSID indicated in the upper-left corner. The rest of the columns show different SST diagnostics. The last panel of the third row shows the magnetic field in the line of sight, obtained from a Milne-Eddington inversion of the \ion{Fe}{I}~6173 line. The dashed red line in the images in the second column marks the location of the IRIS slit in the SJI image to the left. The dashed purple lines in the images in the third column mark the area covered by the IRIS raster. The SST images are down-scaled to the IRIS plate scale.}
\label{fig:all_obs_snapshots}
\end{figure*}

Figure~\ref{fig:all_obs_snapshots} shows images of different IRIS and SST diagnostics of the observations used in this work. Table~\ref{table:observations} summarizes the most relevant parameters of each observation.

\begin{table}[]
\caption{Summary of coordinated IRIS and SST observations. }
\centering
\begin{tabular}{cccccccc}

\hline\hline
Date$^{(a)}$ & Time$^{(b)}$ & OBSID & 
     &  Raster
     & 

 & $\mu ^{(c)}$  & SST overlap$^{(d)}$ \\ \cline{4-6}
& & & \multicolumn{1}{l}{FOV[$\arcsec$]}

&  Cadence [s] & Pointing (solar $X,Y$[$\arcsec$]) 

\\
\hline
20220622 & 072941 & 3620507135 & 5 x 60 & 83 & 381, 228 & 0.88 & 02:37:13 \\ 
20220628 & 072957 & 3620506835 & 5 x 60 & 83 & 216, $-219$ & 0.94 & 03:16:46 \\ 
20230727 & 072953 & 3620506835 & 5 x 60 & 83 & $-349$, $-248$ & 0.90 & 01:05:08 \\ 
20240521 & 073000 & 3620506836 & 5 x 119 & 84 & $-219$, $-139$ & 0.96 & 00:33:30 \\ \hline
\end{tabular}
\tablefoot{$^{(a)}$Observing date in format year, month, day. $^{(b)}$Starting time (UT) in format hour, min, s. $^{(c)}$Cosine of the heliocentric angle ($\mu=\cos\theta$).$^{(d)}$Time duration in format hour, min, s of IRIS - SST overlap.}
\label{table:observations}
\end{table}

The data can be accessed through the IRIS database, available through the public web portal at the IRIS web pages at LMSAL\footnote{\url{https://iris.lmsal.com/search/}}.
Details about the alignment process and the level3 data format can be found in \citet{2020A&A...641A.146R}.
The IRIS data were acquired with 4~s exposure time and a pixel sampling of 0\farcs166. 
    
\section{Inversion's uncertainty computation}\label{appendix:inversion_uncertainty}
\label{appendix:inversion_uncertainty}

\begin{figure*}[!h]
% \centering

\includegraphics[width=1\linewidth]{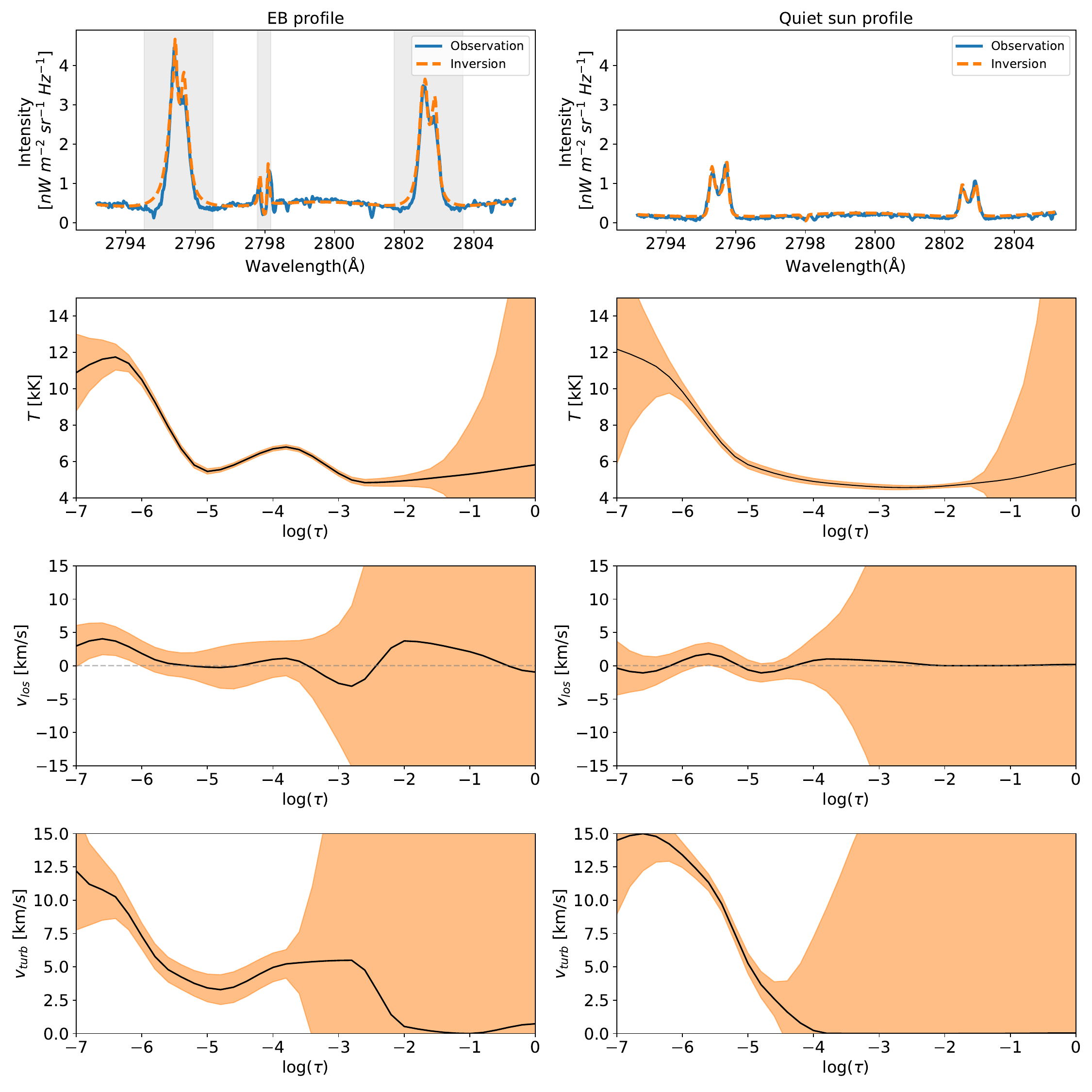}
\caption{Representative inversions using \I2. The left and right columns show the results for an EB and for a quiet-Sun reference profile. First row: Observed spectral profiles (solid blue lines) and synthetic profiles (orange dashed lines). Second, third, and fourth rows: inferred atmospheric stratification for temperature, \vlos, and \vturb. Orange shaded regions indicate 1$\sigma$ uncertainties from the inversion process.}

\label{fig:uncertainty}
\end{figure*}

Figure~\ref{fig:uncertainty} shows an example of two inferences obtained using \I2 and their computed uncertainties for temperature,  the line-of-sight velocity (\vlos), and microturbulence velocity (\vturb). 
Left and right columns correspond to a particularly strong EB profile and to an average quiet sun profile, respectively. 
All the EBs atmosphere models inferred during this work show typical uncertainties contained between both cases.
From the second row of Fig.~\ref{fig:uncertainty}, we see that we can trust the temperature in the range between $\ltau=-6$ and $\ltau=-1.7$. Out of this interval, the uncertainty increases greatly. For the \vlos\ and \vturb\ (third and fourth columns respectively), the interval of higher confidence is shorter, spanning from $\ltau=-6.5$ to $\ltau=-4.5$.

\section{Detected Ellerman bombs}\label{appendix:detected_ebs}

A summary of the 18 detected EBs through all the observations is shown in Table~\ref{table:eb_information}.

\begin{table}[h]
\caption{Summary of detected Ellerman bombs. }
\centering
\begin{tabular}{c|cccccc}
\hline\hline

Date &EB id & \begin{tabular}[c]{@{}c@{}}Lifetime\\ {[}min{]}\end{tabular} & \begin{tabular}[c]{@{}c@{}}Max. size\\ {[}arcsec$^2${]}\end{tabular} & \begin{tabular}[c]{@{}c@{}}First frame$^{(a)}$\\ {[}raster{] }\end{tabular}& \begin{tabular}[c]{@{}c@{}}Last frame$^{(a)}$\\ {[}raster{] }\end{tabular} & 
\begin{tabular}[c]{@{}c@{}}Centroid (X,Y)$^{(a)}$\\ {[}raster pixels{] }\end{tabular}\\
\hline

         &  1 & 20.87 & 1.16  & 23  & 38 & 12  , 265 \\
         &  2 & 8.35 & 2.03  & 23  & 29  & 14  , 241 \\
         &  3 & 8.35 & 2.96  & 51  & 57  & 11  , 264 \\
         &  4 & 23.65 & 1.28  & 82  & 99 & 13  , 263 \\
         &  5 & 6.96 & 0.64  & 97  & 102 & 4   , 281 \\
2022/06/22   &  6 & 27.83 & 1.80  & 106     & 126 & 14, 259 \\
         &  7 & 20.87 & 1.45  & 115  & 130  & 10 ,245 \\
         &  8 & 12.52 & 0.70  & 103  & 112  & 15, 275 \\
         &  9 & 15.30 & 0.87  & 115  & 126  & 4 , 284 \\
         & 10 & 9.74 & 1.28  & 128  & 135   & 14 , 284 \\
         & 11 & 13.91 & 1.92  & 92  & 102   & 13 , 140 \\
         & 12 & 8.35 & 1.34  & 115  & 121   & 13 , 138 \\ \hline
2022/06/28 &  1 & 29.10 & 1.22  & 74& 95    & 12 , 47 \\
            &  2 & 13.86 & 1.10     & 119& 129 & 12, 104 \\ \hline
         &  1 & 22.17 & 2.09  & 44  & 60 &      7   , 267 \\
2023/07/27 &  2 & 18.01 & 0.87  & 77  & 90 &    6   , 235 \\
        &  3 & 11.09 & 0.41  & 75  & 83 &    5   , 270 \\ \hline
2024/05/21 &  1 & 11.17 & 0.99  & 62  & 70 & 9   , 357 \\ \hline \hline
Median   &   - &  13.88 & 1.25 & - & - & - \\ \hline

\end{tabular}
\tablefoot{Summary of detected EBs. $^{(a)}$The information is provided in IRIS level~2 data products format, i.e, the pixels and frames indicated correspond to pixels in the raster and the number of raster.

}
\label{table:eb_information}
\end{table}

\section{Extra EB example}

\begin{figure*}[h!]
    \includegraphics[width=1\linewidth]
    {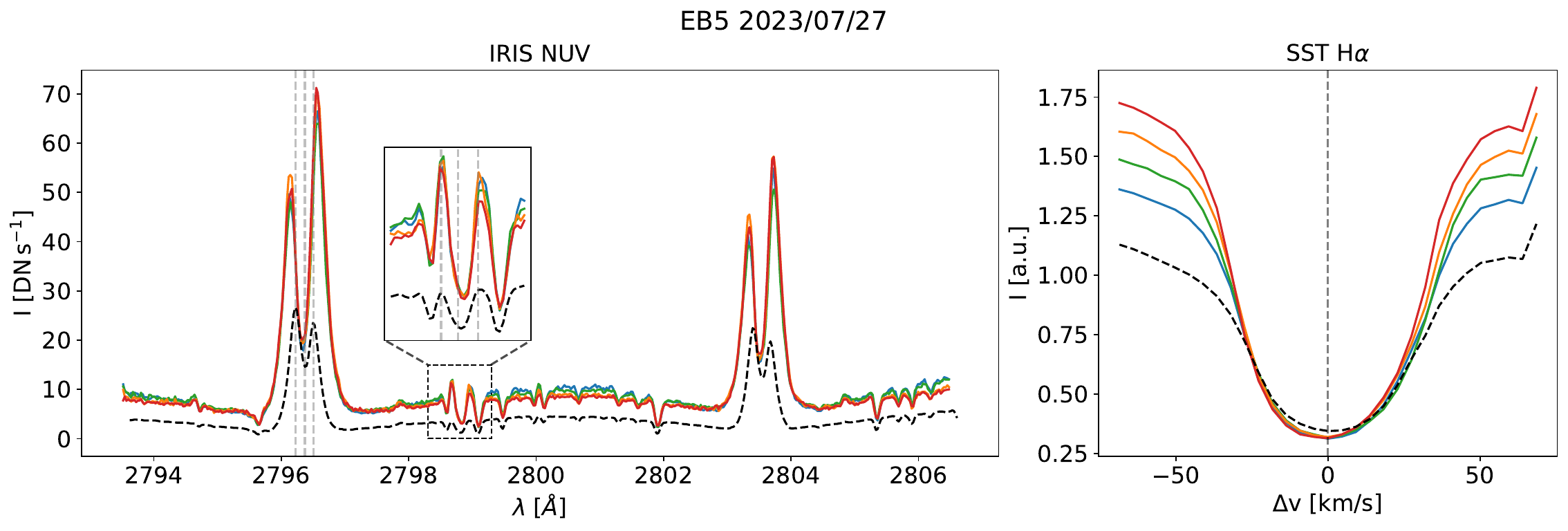}
    \caption{Similar to Fig.~\ref{fig:EB1_2024_05_21} but for EB~5 2022/06/22. This example shows an EB with a strong emission of the \mgii\ ~2$_{v/r}$ peaks but none on the central core. The triplet wings present a moderated enhancement, slightly higher than the bump between \mgii and a relatively strong and pronounced bump between \mgii.}
    \label{fig:EB5_2022_06_22}
\end{figure*}

We show the EB~5 from 2022/06/22 observation in Fig.~\ref{fig:EB5_2022_06_22} as an example of the second trend observed for the \mgii\ lines, explained in Sect.~\ref{sect:summary_spectral_features}. Unlike the previous two EB shown in the manuscript, the EB presented in Fig.~\ref{fig:EB5_2022_06_22} displays enhanced \mgii\ ~2$_{v/r}$ peaks while the line cores are very close to the reference profile.

\section{EB statistics in DN~s$^{-1}$}\label{appendix:stats_DNs}

\begin{figure}[h]
    \centering
    \includegraphics[width=0.90\linewidth]{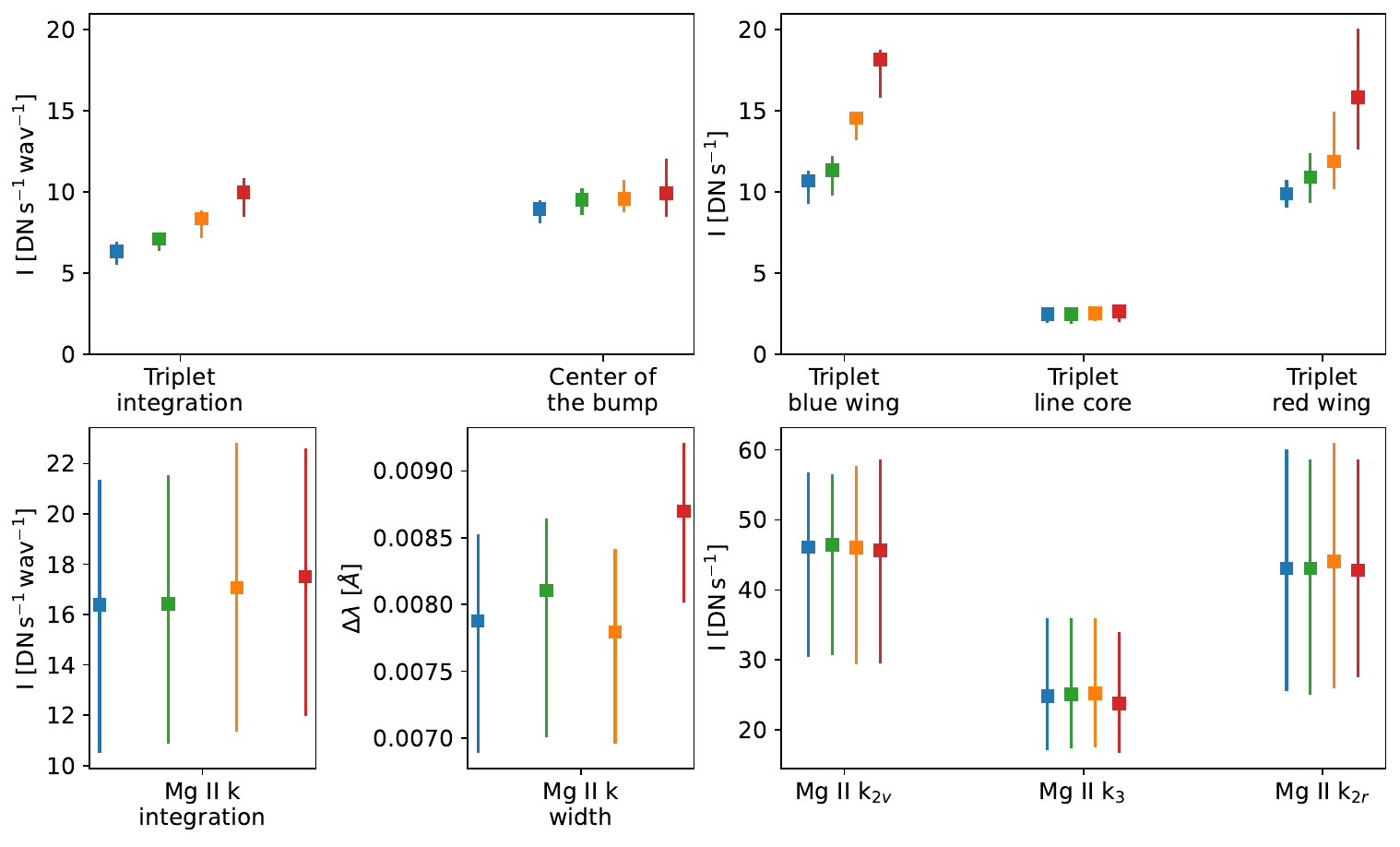}
    \caption{Statistical summary of the IRIS NUV spectral features for all the detected EBs (in DN~s$^{-1}$). The layout and percentile definitions are identical to Fig.~\ref{fig:gen_stats}, but represented here in physical Data Number per second (DN~s$^{-1}$) units rather than relative enhancements.}
    \label{fig:gen_stats_dns}
\end{figure}

Figure~\ref{fig:gen_stats_dns} shows the same information as Fig.~\ref{fig:gen_stats} but with DN~s$^{-1}$ units. This is provided so it is possible to compare our findings with any other IRIS observations. 

    \twocolumn
\end{appendix}

\appendix

\end{document}